\documentclass[12pt]{iopart}
\usepackage{iopams}  
\usepackage{graphicx}
\usepackage{array}
\usepackage{dcolumn}
\usepackage{bm}
\usepackage{float}
\usepackage{enumitem}
\usepackage[colorinlistoftodos, textwidth=4cm, shadow]{todonotes}

\usepackage{dcolumn}

\newcommand{\degrees}{$^{\circ}$}

\newcommand{\FAS}{Fe$_{5-x}$AsTe$_2$\,}

\newcommand{\FASx}{Fe$_{5-x}$AsTe$_2$\,}

\newcommand{\TN}{$T_{\rm N}$\,}

\renewcommand{\theequation}{\textbf{\arabic{equation}}}
\renewcommand{\thefigure}{\textbf{\arabic{figure}}}
\renewcommand{\thetable}{\textbf{\arabic{table}}}
\renewcommand{\figurename}{\textbf{Figure}}
\renewcommand{\tablename}{\textbf{Table}}

\begin{document}



\title{Tuning the Room Temperature Ferromagnetism in Fe$_{5}$GeTe$_2$ by Arsenic Substitution}

\author{Andrew F. May,* Jiaqiang Yan, Raphael Hermann, Mao-Hua Du, and Michael A. McGuire}
\address{Materials Science and Technology Division, Oak Ridge National Lab, Oak Ridge, TN 37831}
\ead{mayaf@ornl.gov}
\vspace{10pt}
\begin{indented}
\item[]October 21, 2021
\end{indented}

\begin{abstract}

In order to tune the magnetic properties of the cleavable high-Curie temperature ferromagnet Fe$_{5-x}$GeTe$_2$, the effect of increasing the electron count through arsenic substitution has been investigated.  Small additions of arsenic (2.5 and 5\%) seemingly enhance ferromagnetic order in polycrystalline samples by quenching fluctuations on one of the three magnetic sublattices, whereas larger As concentrations decrease the ferromagnetic Curie temperature ($T_{\rm C}$\,) and saturation magnetization.  This work also describes the growth and characterization of Fe$_{4.8}$AsTe$_2$ single crystals that are structurally analogous to Fe$_{5-x}$GeTe$_2$ but with some phase stability complications.  Magnetization measurements reveal dominant antiferromagnetic behavior in Fe$_{4.8}$AsTe$_2$ with a N\'{e}el temperature of $T_{\rm N}$ $\approx$42\,K. A field-induced spin-flop below $T_{\rm N}$ results in a switch from negative to positive magnetoresistance, with significant hysteresis causing butterfly-shaped resistance loops.  In addition to reporting the properties of Fe$_{4.8}$AsTe$_2$, this work shows the importance of manipulating the individual magnetic sublattices in Fe$_{5-x}$GeTe$_2$ and motivates further efforts to control the magnetic properties in related materials by fine tuning of the Fermi energy or crystal chemistry.  

\end{abstract}

%
\vspace{2pc}
%
\submitto{\TDM; online at http://iopscience.iop.org/article/10.1088/2053-1583/ac34d9}
%
%
\ioptwocol
%


\section{Introduction}

Itinerant van der Waals (vdW) magnets provide promising platforms to study the complex relationships between emergent magnetic phenomena and crystal chemistry. Studies of magnetism in layered vdW materials probe the nature of magnetic order and interactions in the bulk and 2D limit, how these fundamental properties can be tuned, and the various types of device-related responses that may emerge in the pure system or via heterostructure design.\cite{Burch2018,Wang2020rev,Huang2020rev}   Of the pertinent vdW materials, several metallic Fe-Ge-Te phases possess some of the highest magnetic ordering temperatures.\cite{Deiseroth2006,Stahl2018,May2019acs,Jothi2019}  Ferromagnetism in monolayer Fe$_{3-x}$GeTe$_2$ has been demonstrated, and of particular interest is the increase in the Curie temperature $T_{\rm C}$ from $\approx$100\,K to over 300\,K caused by electrochemical gating of few-layer flakes, perhaps due to the intercalation of lithium.\cite{Fei2018,Deng2018} Importantly, $T_{\rm C}$ of bulk Fe$_{5-x}$GeTe$_2$ ranges from 270-310\,K and can be further enhanced by cobalt or nickel substitution.\cite{Stahl2018,May2019acs,May2019,Tian2020,May2020}  Fe$_{5-x}$GeTe$_2$ and related compositions are thus prime candidates for incorporation into synthetic vdW heterostructures, where the properties can be tuned by the local composition.  For instance, the stabilization of longer-range emergent phenomena such as topological spin textures (e.g. skyrmions) is being heavily pursued in Fe$_{3-x}$GeTe$_2$ and related vdW heterostructures.\cite{Ding2020,wang2020characteristics,wu2020neel,yang2020creation,Park2021skrymion}  In general, Fe$_{5-x}$GeTe$_2$ and related Fe$_{4}$GeTe$_2$ have garnered significant attention recently due to their high ordering temperature and complex behaviors.\cite{Wu2021,Yang2021,Yamagami2021,Li2020magnetic,Tan2021gate,Ly2021direct,Ohta2021butterfly,Zhang2020,Seo2020} The present work was motivated by identifying further routes to tune the magnetism in Fe$_{5-x}$GeTe$_2$ and related bulk phases so that novel properties and logical designs of heterostructures can be achieved.

Fe$_{5-x}$GeTe$_2$ contains significant disorder associated with a large concentration of vacancies on one of three Fe sublattices (see Fig.\,\ref{XRD}a).\cite{May2019acs} A large amount of stacking disorder also exists within the crystals.\cite{Stahl2018,May2019}  Control over Fe content has not been demonstrated, but the magnetic properties of bulk and thin film Fe$_{5-x}$GeTe$_2$ depend on thermal processing and differences between polycrystalline and single crystalline samples have been observed.\cite{Stahl2018,May2019acs,May2019,Ohta2020} M\"{o}ssbauer spectroscopy has evidenced that spin fluctuations on the (atomically disordered) Fe1 sublattice persist to $\approx$100\,K despite magnetic ordering on the other sublattices near 300\,K.\cite{May2019acs}   Recently, evidence linking magnetic order on the Fe1 sublattice with a competing charge order state has also been discussed,\cite{Wu2021} and inversion breaking associated with the atomic (vacancy) ordering on the Fe1(Ge) sublattice has been proposed as a source for helimagnetic order.\cite{Ly2021direct} The ordering of moments on the Fe1 sublattice impacts the electrical properties, the lattice parameters, and the magnetic anisotropy,\cite{May2019acs,May2019} and therefore controlling magnetism on the Fe1 sublattice is essential for tuning the properties of Fe$_{4.8}$GeTe$_2$. For instance, stronger inter-sublattice coupling may result in stronger magnetism and this could explain why Ni or Co substitution in Fe$_{5-x}$GeTe$_2$ increases $T_{\rm C}$.\cite{Stahl2018,May2020}  By contrast, Ni or Co substitution into Fe$_{3-x}$GeTe$_2$ suppresses $T_{\rm C}$,\cite{Drachuck2018,Tian2019} as does decreasing the Fe content or substituting As for Ge.\cite{Verchenko2015,May2016,Yuan2017} 

Due to the unique response of Fe$_{5-x}$GeTe$_2$ to transition metal substitutions, we were motivated to investigate the impact of As substitution for Ge in Fe$_{5}$Ge$_{1-y}$As$_y$Te$_2$. Investigation of polycrystalline Fe$_{5}$Ge$_{1-y}$As$_y$Te$_2$ samples reveal an enhancement in $T_{\rm C}$ for low As contents ($y$=0.025 and 0.05), but a clear decrease in $T_{\rm C}$ and the saturation moment were observed for 0.25 $<$ $y$ $<$ 0.75.  These results suggest that fine tuning of the crystal chemistry in Fe$_{5-x}$GeTe$_2$ is a viable means of controlling its room temperature magnetic properties.  We also report the crystal structure and physical properties of Fe$_{5-x}$AsTe$_2$ ($x \approx$ 0.2), which contains significant disorder.  Magnetization measurements support a canted antiferromagnetic ground state in Fe$_{4.8}$AsTe$_2$, and butterfly-shaped magnetoresistance loops are found to be driven by a hysteretic meta-magnetic transition.

\section{Results and Discussion}

\subsection{Structural Characterization}

We utilized polycrystalline samples of Fe$_{5}$Ge$_{1-y}$As$_y$Te$_2$ to examine how the lattice and magnetism evolves with arsenic substitution, though we note that structural complexities of the Fe$_{5-x}$GeTe$_2$ system could drive subtle differences in the magnetic behavior of polycrystalline versus single crystalline samples.   The  samples were quenched from 750\degrees C to promote chemical homogeneity, and x-ray powder diffraction data were collected at ambient conditions.  A Le Bail fitting procedure was used to extract the lattice parameters because significant anisotropic peak broadening attributed to stacking disorder (intrinsic or induced by sample preparation) hinders Rietveld refinement of the diffraction data; the rhombohedral Fe$_{5-x}$GeTe$_2$ lattice symmetry was utilized.  The powder diffraction patterns are shown in the Supporting Materials.\cite{Supporting}   The presence of As was confirmed by energy dispersive spectroscopy (Bruker TM3000 with Quantax EDS at 15 keV) performed for the polycrystalline Fe$_{5}$Ge$_{1-y}$As$_y$Te$_2$ samples. The As/Ge L-series peak overlaps were found to complicate the accurate measurements of the Ge/As ratio, especially at low concentrations. The measurements were found to overestimate the As content, as demonstrated by measurements on an As-free crystal that indicated an artificial As content up to $\approx$ 5\% As (relative to Ge). Measurements on samples with nominal As contents of 25, 50, and 75\% returned 29(1), 53(1), and 77(1)\% relative to Ge, showing that the actual and nominal concentrations are similar. The nominal value of $y$ is used throughout the text to establish the qualitative trends with arsenic substitution.  

\begin{figure}[ht!]%
\includegraphics[width=1.05\columnwidth]{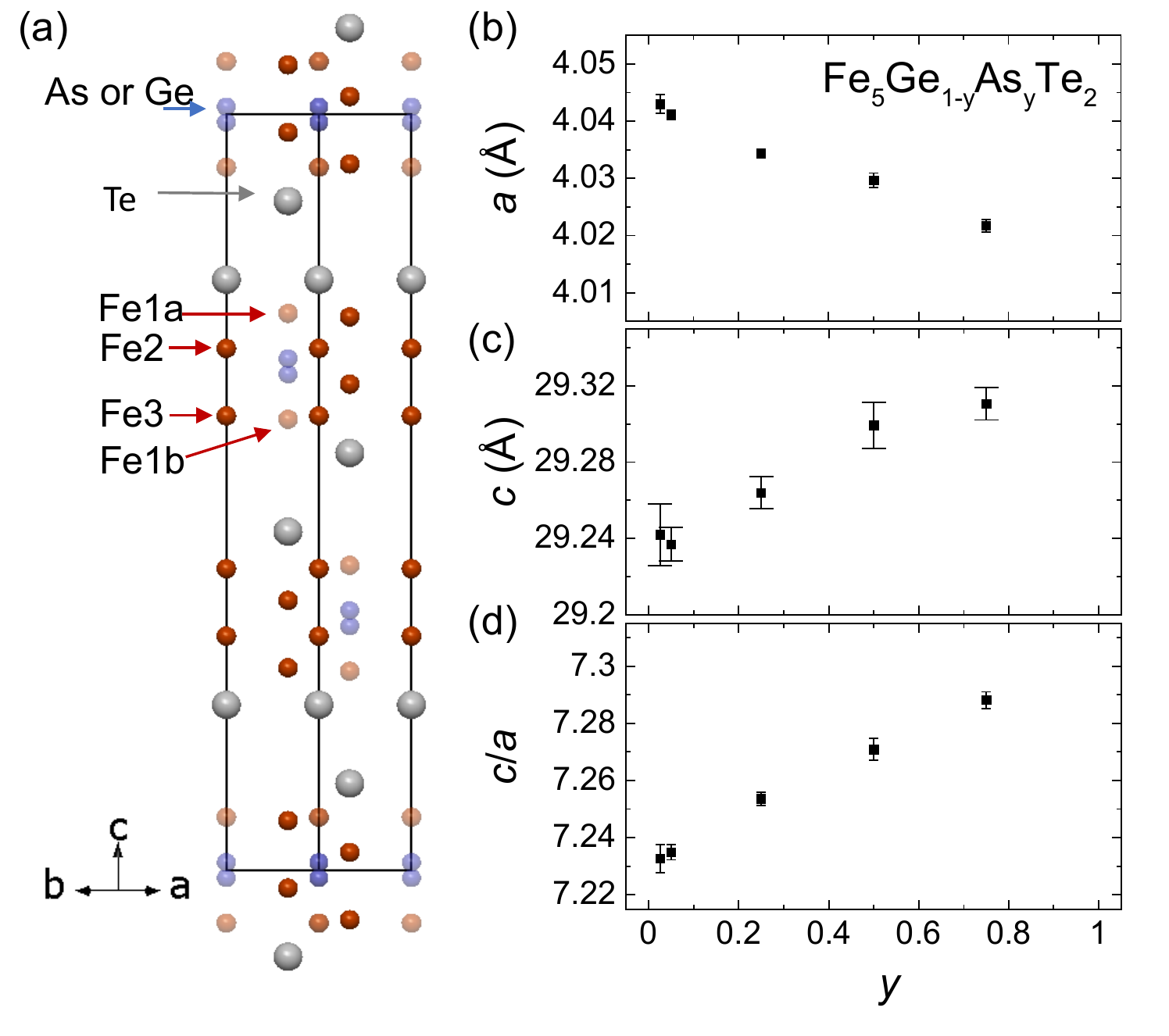}%
\caption{(a) Crystal structure of Fe$_{5}$(Ge,As)Te$_2$ with partially transparent Fe1 and As positions indicating split-sites with up to 50\% occupancy. (b,c) Lattice parameters obtained by fitting room temperature x-ray diffraction data for quenched polycrystalline samples and  (d) the ratio of lattice parameters.}%
\label{XRD}
\end{figure}

As shown in Fig. \ref{XRD}(b-d), the substitution of As for Ge leads to a contraction of the unit cell within the $\textit{ab}$-plane and an expansion along the $c$-axis.  This results in an increase in the ratio $c/a$, potentially implying an increase in the 2D character with increasing arsenic content. Similar lattice trends were observed when As was substituted for Ge in Fe$_{3-x}$GeTe$_2$.\cite{Yuan2017}  In Fe$_{5-x}$GeTe$_2$, the $h0l$ diffraction peaks are significantly broadened when samples are cooled slowly because there is a structural transition near $\approx$570\,K that induces stacking disorder upon cooling; however, quenching results in a metastable state where sharp $h0l$ reflections are maintained.\cite{May2019}  In the mixed Ge-As samples, broadening of diffraction peaks due to stacking faults was found to decrease with increasing As content from 2.5 to 75\% arsenic relative to Ge, though some broadening is observed despite quenching (see Supporting Materials).  The trends observed in Fig.\,\ref{XRD}(b,c,d) are rather robust despite peak broadening because the 00$l$ and $hhl$ peaks are typically not broadened by the stacking disorder.\cite{May2020practical}  

Single crystals of Fe$_{4.8}$AsTe$_2$ were grown in the presence of iodine as discussed in the Methods section.  Firstly, we note that the crystals are plate-like in nature and behave similar to Fe$_{5-x}$GeTe$_2$ during simple cleaving tests using adhesive tapes.  Since exfoliation of Fe$_{5-x}$GeTe$_2$ to near monolayer limit has been demonstrated,\cite{May2019acs,Ohta2020,Tan2021gate} it seems likely that similar exfoliation of the arsenide or arsenic-containing crystals would be possible. Dedicated efforts are necessary to examine this characteristic in detail, and such endeavors should probably treat Fe$_{5-x}$AsTe$_2$ flakes as air sensitive.   To address the composition of the crystals, wavelength dispersive spectroscopy was performed in a JEOL electron microprobe on the as-grown facets of slow cooled crystals, and this chemical analysis technique yielded an average composition of Fe$_{4.77(7)}$As$_{0.97(2)}$Te$_{2.00(5)}$.  We thus refer to the crystals using the composition Fe$_{4.8}$AsTe$_2$ for simplicity.

Samples of the arsenic end-member Fe$_{4.8}$AsTe$_2$ generally possess a large degree of stacking disorder as well as a secondary phase. For slow cooled crystals, the primary phase has $c$=29.51(2)\AA\, and the secondary phase has a small cell with $c$=28.67(1)\AA\, as obtained by fitting the 00$l$ Bragg reflections from a diffraction measurement off a crystal facet.  These values are notably different from those in the polycrystalline As-Ge alloys.  While the larger $c$-axis parameter of the main phase could somehow relate to the stacking disorder, the significantly reduced $c$-axis parameter of the secondary phase likely has a chemical or structural origin.   The extent to which the smaller-cell phase is present appears to depend on fine details of the synthesis and some related data are shown in the Supporting Materials; further investigations into the phase stability and local homogeneity of Fe$_{5-x}$AsTe$_2$ are necessary.   As discussed below, the act of cooling slowly from the growth temperature promotes long-range magnetic order in Fe$_{4.8}$AsTe$_2$ crystals whereas thermal quenching appears to induce glassy magnetic behavior. As such, this article focuses on Fe$_{4.8}$AsTe$_2$ crystals that are cooled in the furnace from the growth temperature over 8-12\,h.  

Single crystal x-ray diffraction data reveal that the slow-cooled Fe$_{4.8}$AsTe$_2$ crystals contain significant stacking disorder or local variations of $c$, which precludes structural determination from diffraction data.  Less stacking disorder was observed in a quenched crystal for which the single crystal x-ray diffraction data were able to be refined using the Fe$_{5-x}$GeTe$_2$ crystal structure with $a$ = 4.0088(6)\AA\, and $c$ = 29.279(6)\AA\, at $T$ = 220\,K (see Supporting Materials for a comparison of the data).  These results suggest that Fe$_{5-x}$AsTe$_2$ and Fe$_{5-x}$GeTe$_2$ have similar structural units as shown in Fig.\,\ref{XRD}(a).  Fe$_{5-x}$AsTe$_2$ has a complex temperature-dependent phase evolution that promotes phase separation via changes in layer stacking, composition, and/or site occupancy; further work is needed to understand the phase stability.  Importantly, a $\sqrt{3}a \times \sqrt{3}a$ supercell that was observed in Fe$_{5-x}$GeTe$_2$ was also observed for Fe$_{4.8}$AsTe$_2$ by single crystal x-ray diffraction (both quenched and furnace-cooled crystals).  This further supports the structural similarity between the two materials because this in-plane supercell is related to occupancy on the Fe1a,b sublattice (a unique structural feature).  Short correlation lengths along [001] preclude structural solution from x-ray diffraction data and cause a streaking of diffraction intensity (see Supporting Materials).

\subsection{Physical Properties}

\begin{figure*}[ht!]%
\includegraphics[width=2.05\columnwidth]{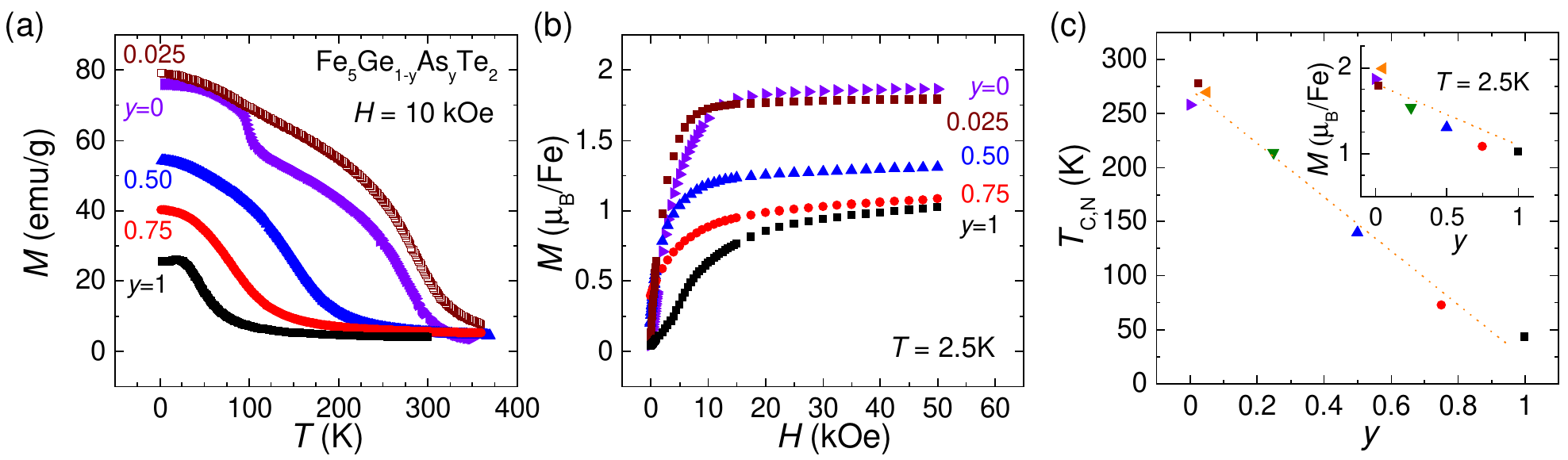}%
\caption{ Magnetization data for polycrystalline Fe$_{5}$Ge$_{1-y}$As$_y$Te$_2$ samples.  (a) Temperature-dependent magnetization and (b) isothermal magnetization data. (c) Magnetic ordering temperatures obtained using magnetization data collected in a small applied field of $H$ = 100\,kOe as discussed in the Supporting Materials. The inset shows the saturation magnetization reached versus nominal arsenic content $y$ ($T$=2.5\,K, $H$=50\,kOe). Antiferromagnetic behavior is observed for the Fe$_{5}$AsTe$_2$ sample while the mixed As/Ge alloy compositions display ferromagnetic character.}%
\label{MagMain}
\end{figure*}

We first discuss how arsenic substitution for germanium changes the properties in polycrystalline Fe$_{5}$Ge$_{1-y}$As$_y$Te$_2$ and then consider properties of single crystal Fe$_{4.8}$AsTe$_2$. Our primary interest is in the evolution of the magnetic properties, and magnetization ($M$) measurements are the key characterization technique utilized.   Figure \ref{MagMain}(a) contains the temperature-dependent $M$ data for polycrystalline Fe$_{5}$Ge$_{1-y}$As$_y$Te$_2$ at various compositions and Fig.\ref{MagMain}(b) plots the field-dependent $M$ data at $T$=2.5\,K.  The overall trend is for decreasing ordering temperature and induced (saturation) moment with increasing arsenic content, as summarized in Fig.\ref{MagMain}(c).  However, close inspection reveals a more complicated scenario with different behavior observed for small arsenic concentrations  ($y$=0.025 and 0.05) and a comparison to the polycrystalline $y$=0 data is important for understanding these results.  In the parent material Fe$_{5-x}$GeTe$_2$, the magnetic behavior of polycrystalline samples is slightly different than that in the single crystals, though the main features appear consistent between the two. However, the Curie temperature of powders appears to be slightly lower than that of crystals and the first-order magnetostructural transition is only observed in the quenched and metastable crystals.\cite{Stahl2018,May2019acs,May2019}  As such, it is important to compare properties of the arsenic containing powders to the polycrystalline samples of the parent material. We observed plate-like morphology of the crystallites formed during the annealing of polycrystalline samples, and thus it seems reasonable to speculate that single crystal growth of Fe$_{5}$Ge$_{1-y}$As$_y$Te$_2$ is possible for at least certain compounds. However, detailed growth studies are necessary to evaluate how crystal growth impacts the chemistry and properties of such samples.

The $M$($T$) curve for Fe$_{5-x}$GeTe$_2$ ($y$=0) in Fig.\,\ref{MagMain}a is characterized by a strong rise in $M$($T$) upon cooling below $T_{\rm C}$ $\approx$ 258\,K and there is a strong anomaly near 100\,K.  The signature in the magnetization near 100\,K has been associated with the establishment of magnetic order on the Fe1 sublattice (observed in both powders and crystals).\cite{May2019acs}  The magnetism on the Fe1 sublattice impacts many properties and therefore controlling the Fe1 sublattice is a primary route for manipulating the magnetism in Fe$_{5-x}$GeTe$_2$.  For instance, there is a coupled magnetoelastic response and ordering of the Fe1 moments impacts the electronic properties and magnetic anisotropy,\cite{May2019acs,May2019} with easy-axis [001] anisotropy strengthening upon cooling below 100\,K.  Due to this evolution of the anisotropy, topological vortex features known as (anti)merons formed at domain walls become unstable at low $T$ in Fe$_{5-x}$GeTe$_2$.\cite{Gao2020} The existence of a competing charge order above 100\,K has also been discussed recently.\cite{Wu2021}

\begin{table}[]
\caption{Density functional theory results of total ferromagnetic (FM) moments on different Fe sublattices for Fe$_5$AsTe$_2$ and Fe$_5$GeTe$_2$.  Due to the simulated supercell with a checkerboard occupation of the Fe1 sublattice, atoms at different Fe2 and Fe3 positions have different local environments and moments. The data for Fe$_5$AsTe$_2$ were obtained using crystallographic parameters obtained from a quenched single crystal.}
\label{DFT}
\begin{tabular}{|c|c|c|c|c|}
\hline
     & \multicolumn{2}{c|}{Fe$_5$AsTe$_2$}  & \multicolumn{2}{c|}{Fe$_5$GeTe$_2$}   \\ \hline
FM moment & \multicolumn{2}{c|}{9.29/f.u.} & \multicolumn{2}{c|}{10.92/f.u.} \\ \hline
sites           & \multicolumn{4}{c|}{FM Moment ($\mu_B$/Fe)}                      \\ \hline
Fe1             & 1.19           &               & 1.98           &                \\ \hline
Fe2             & 1.99           & 1.09          & 2.14           & 1.96           \\ \hline
Fe3             & 2.64           & 2.38          & 2.52           & 2.32           \\ \hline
\hline
\end{tabular}
\end{table}

The magnetic anomaly associated with the formation of magnetic order on the Fe1 sublattice is notably absent in the magnetization data for the polycrystalline arsenic-containing samples.  This is true for all arsenic containing samples that were measured (including crystals of the pure arsenide end member).  This suggests that arsenic substitution quenches spin fluctuations on the Fe1 sublattice and leads to an enhanced Curie temperature for powders with $y$=0.025 and 0.05.  These results show that magnetism on the Fe1 sublattice is very sensitive to small changes in the Fermi level or chemistry in Fe$_{5-x}$GeTe$_2$.   $M$($T$) data provide a quick screening for this behavior, though more local probes like M\"{o}ssbauer spectroscopy\cite{May2019acs} or other zero field measurements are required to conclude if the Fe sublattices are not independent after As substitution for Ge.  Changes to magnetic anisotropy may impact the magnetization measurements and thus hinder a direct conclusion of the underlying behavior, especially in applied fields.  Finally, as noted above, the behavior in single crystals may vary from that in polycrystalline samples due to the metastability and disorder in the Fe$_{5-x}$GeTe$_2$ family.

In addition to the change in behavior of $M$($T$) data, the isothermal magnetization $M$($H$) in  Fig.\,\ref{MagMain}c reveal a decreased critical field for saturation in the $y$=0.025 sample in comparison to the $y$=0 sample.  This `softening' of the magnetism implies a loss of magnetic anisotropy, which is also consistent with arsenic substitution impacting magnetism on the Fe1  sublattice.  A change in the anisotropy was also observed in cobalt substituted Fe$_{5-y}$Co$_y$GeTe$_2$, where crystals with $y\approx$1 had easy-plane anisotropy that was opposite to the easy-axis [001] anisotropy of the $y$=0 parent.\cite{Tian2020,May2020}  The control over magnetic anisotropy, both sign and magnitude, is important because it has implications for the design of heterostructures where the anisotropy and magnetic properties are tuned locally, and because anisotropy is considered an important ingredient to form magnetic order in the 2D limit.  Indeed, magnetic anisotropy is an important parameter for the stabilization of topological spin textures such as skyrmions.

The saturation moment is reduced by increased arsenic content, as illustrated in the inset of Fig.\ref{MagMain}(c).  The rather continuous decrease in the saturation moment could be linked to the smooth evolution of the lattice parameters (Fig.\ref{XRD}), as was suggested for the behavior in Fe$_{3-y}$Ge$_{1-x}$As$_x$Te$_2$.\cite{Yuan2017}  Our density functional theory (DFT) calculations do not predict a strong decrease in the net moment of idealized compositions Fe$_{5}$AsTe$_2$ relative to Fe$_{5}$GeTe$_2$.  For instance, as summarized in Table \ref{DFT}, our DFT results predict an average ferromagnetic moment of 1.84$\mu_B$/Fe in Fe$_{5}$AsTe$_2$ compared to 2.15 $\mu_B$/Fe in Fe$_{5}$GeTe$_2$. The DFT results suggest that the Fe1 sublattice (and neighboring Fe2) are the most impacted by the presence of As.  While DFT suggests a dominant FM ground state in atomically-ordered Fe$_5$AsTe$_2$, an AFM order with AFM coupling along [001] was found to be only $\approx$0.6\,meV/Fe higher in energy for a different (zig-zag) occupancy pattern on the Fe1a,b sublattice.  In Fe$_5$GeTe$_2$, the first competing magnetic order was calculated to be more than 2\,meV/Fe above the ground state and primitive layer stacking was found to decrease the stability of FM relative to AFM in Fe$_5$GeTe$_2$.\cite{May2020}  These trends may help explain the formation of a non-compensated antiferromagnetic structure in Fe$_{4.8}$AsTe$_2$ crystals where significant stacking disorder is evidenced by the diffraction data.  Recently, experiments have demonstrated that gating ultra-thin Fe$_{5-x}$GeTe$_2$ seemingly produces an antiferromagnetic state,\cite{Tan2021gate} and it has been shown that cobalt substitution also induces AFM.\cite{Tian2020,May2020}  Also, calculations have suggested that bilayer Fe$_{5}$GeTe$_2$ will be antiferromagnetic.\cite{Yang2021}  The DFT calculations are performed at an idealized stoichiometry and for idealized structures with specific Fe1a,b occupancy configurations and without any stacking faults. The situation in the crystals is much more complex, and the stacking disorder and the random Fe1 distributions could induce local AFM coupling and this may also reduce the saturation moment. In the limit of strong AFM coupling, the field-induced state that looks like saturation may in fact be a ferrimagnetic spin configuration.  Further exploration of the magnetic moment with a local probe or a high-field measurement are necessary to understand the discrepancy between the induced moment and the theoretical moment in Fe$_{4.8}$AsTe$_2$ samples.

\begin{figure}[h!]%
\includegraphics[width=0.95\columnwidth]{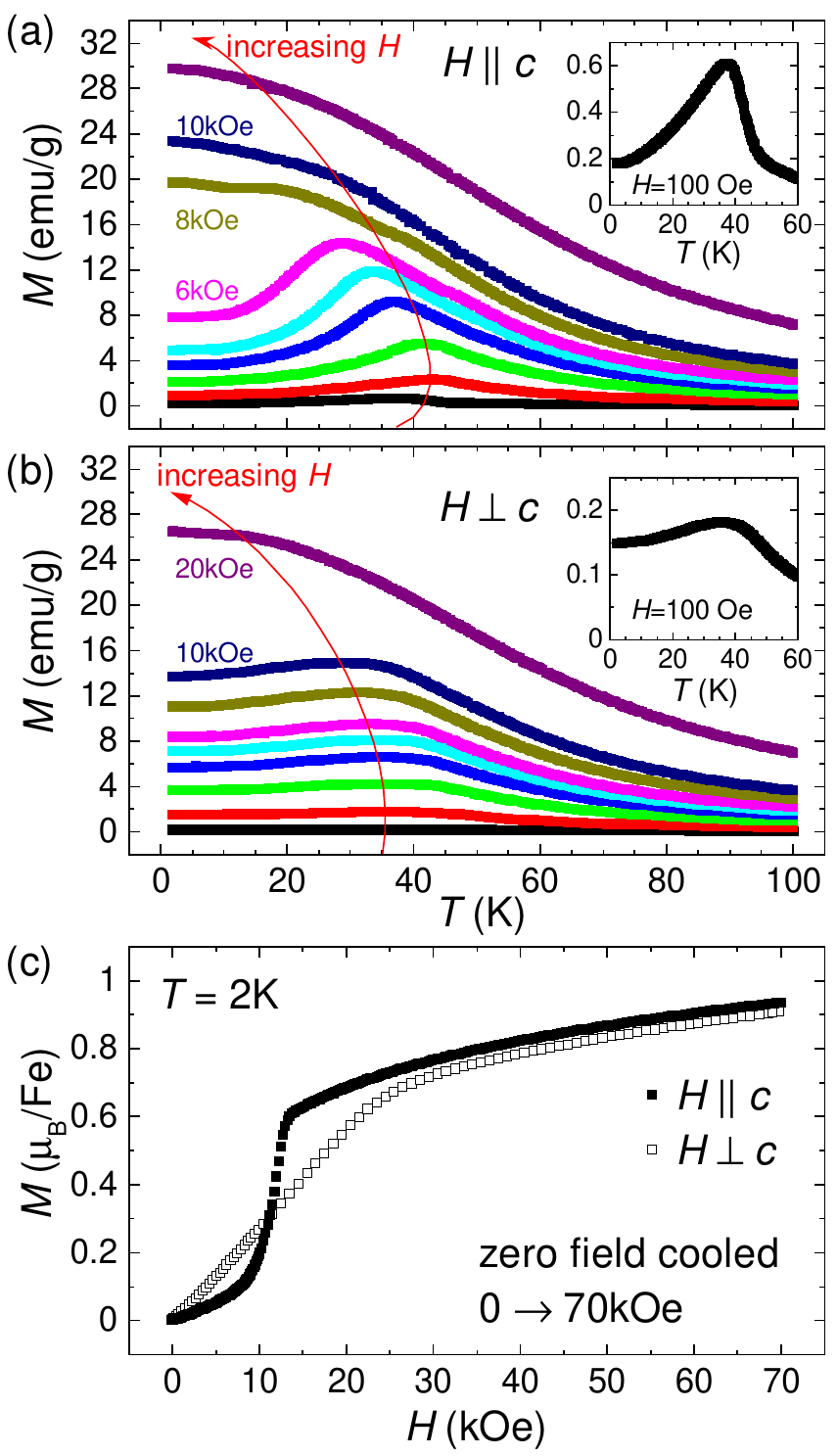}%
\caption{Anisotropic magnetization data for single crystalline Fe$_{4.8}$AsTe$_2$, with temperature-dependent data for (a) $H \parallel c$  and (b) $H \perp c$, and (c) isothermal data after cooling in zero applied field. The insets in (a,b) show the low-field data near $T_N$ using the same vertical axis units (emu/g) as the main panels.  In (a,b), data for applied fields of $H$ = 0.1, 1, 2.5, 4, 5, 6, 8, 10, 20\,kOe are shown.}%
\label{MagXtl}
\end{figure}

We now discuss the magnetic behavior of Fe$_{4.8}$AsTe$_2$ in greater detail. The $M$($T$) data for Fe$_{4.8}$AsTe$_2$ are qualitatively different than those of the ferromagnetic Fe$_{5}$Ge$_{1-y}$As$_y$Te$_2$ samples with $y \leq 0.75$.   Indeed, the $M$($T$) data for Fe$_{4.8}$AsTe$_2$ have a cusp that is characteristic of antiferromagnetic ordering, as shown in Fig.\ref{MagXtl}(a,b) and for the polycrystalline sample in Fig.\ref{MagMain}(a).  We define \TN = 42\,K based on ac magnetic susceptibility data taken in zero applied dc magnetic field (shown in Supporting Materials).  The temperature of the cusp in the dc $M$($T$) data is generally suppressed to lower $T$ with increasing applied field $H$, as expected for AFM order.  However, upon increasing the applied field from $H$ $\parallel$ $c$ = 0.1 to 4\,kOe  there is an increase in the temperature where the cusp is observed, and this qualitative behavior is suggestive of a non-compensated AFM order (a canted AFM order with a weak ferromagnetic component).  An additional view of the data that highlights this behavior is shown in the Supporting Materials.

The isothermal magnetization data presented in Fig.\ref{MagXtl}(c) portray the anisotropic magnetic response of  Fe$_{4.8}$AsTe$_2$ at $T$=2\,K.  Of particular importance here is the apparent spin-flop transition observed when the applied field is parallel to the $c$-axis.  Upon increasing the field from a zero field cooled (ZFC) condition, the spin-flop occurs near 10\,kOe.  This implies that the moments are oriented primarily along the $c$-axis for $H$=0, and they reorient to perpendicular to the [001] direction of the applied field near 10\,kOe before slowly rotating towards a saturated state.  A metamagnetic transition suggesting easy-axis anisotropy was also observed in the AFM phase induced by cobalt doping of Fe$_{5-x}$GeTe$_2$.\cite{Tian2020,May2020}  When the field is applied within the basal plane ($H$ $\perp$ $c$) of Fe$_{4.8}$AsTe$_2$, the magnetization increases continuously up to around 25\,kOe and then gradually increases towards an assumed saturation.  This behavior also supports the hypothesis of long range antiferromagnetic order in these furnace cooled Fe$_{4.8}$AsTe$_2$ crystals.  By contrast, the crystals that were quenched from 750 \degrees C did not display anisotropic $M$($H$) and the magnetic properties were generally consistent with the existence of short range AFM correlations and possible glassy behavior (see Supporting Materials for related data).  Interestingly, the quenched crystals appear to have less stacking disorder and this may suggest that small changes in the Fe1a,b sublattice or lattice strain strongly impact the magnetism.

\begin{figure}[h!]%
\includegraphics[width=0.95\columnwidth]{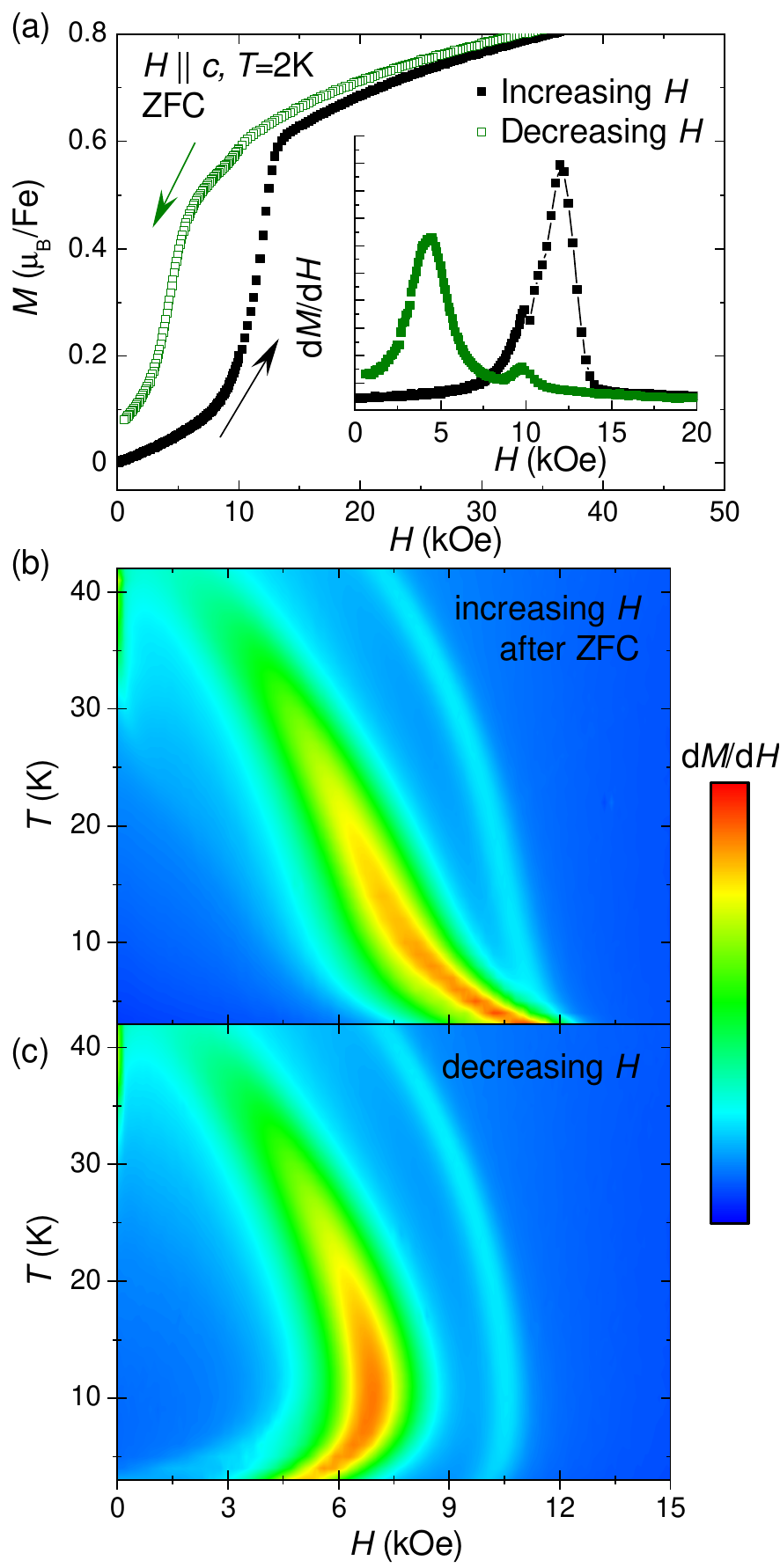}%
\caption{Isothermal magnetization for $H \parallel c$ in furnace cooled single crystals of Fe$_{4.8}$AsTe$_2$.  (a) Data for increasing and decreasing the applied field after zero field cooling (ZFC) with a maximum field of 70\,kOe reached.  The inset shows the derivative d$M$/d$H$.  (b,c) Contour plots of d$M$/d$H$ as a function of $T,H$ for (b) increasing $H$ and (c) decreasing $H$.}%
\label{MHparC}
\end{figure}

The spin-flop transition has significant field hysteresis, as illustrated in Fig.\,\ref{MHparC}(a).  Upon decreasing the field to $H$=0 from high fields, a remanent moment is observed.  A smaller, but finite, remanent moment is also observed for $H$ $\perp$ $c$.  The inset of Fig.\,\ref{MHparC}(a) displays the derivative of the isothermal magnetization d$M$/d$H$ and small features can be observed in addition to the main spin-flop transition.

The critical field of the spin-flop increases upon cooling from $T_{\rm N}$ and so does the hysteresis associated with this meta-magnetic transition. These trends can be inferred from the contour plots in Fig.\,\ref{MHparC}(b,c) where the color scale is related to the value of d$M$/d$H$ and thus the highest intensity (red) relates to the spin-flop transition where increasing (decreasing) the field rapidly increases (decreases) the magnetization.  The data in Fig.\,\ref{MHparC}(b) were obtained while increasing $H$ after cooling in zero field, and thus they demonstrate the increasing anisotropy and critical field upon cooling.   Both the increasing- and decreasing-field data contain shoulders to the spin-flop transition (a second band of large d$M$/d$H$ at higher fields). These shoulders may be caused by complex domain behavior or inhomogeneity in the sample that promote different anisotropy energies.   Similarly, the existence of a remanence could be linked to the large hysteresis of the spin-flop or the presence of complex domain walls that promote a small residual moment.

\begin{figure*}[ht!]%
\includegraphics[width=1.95\columnwidth]{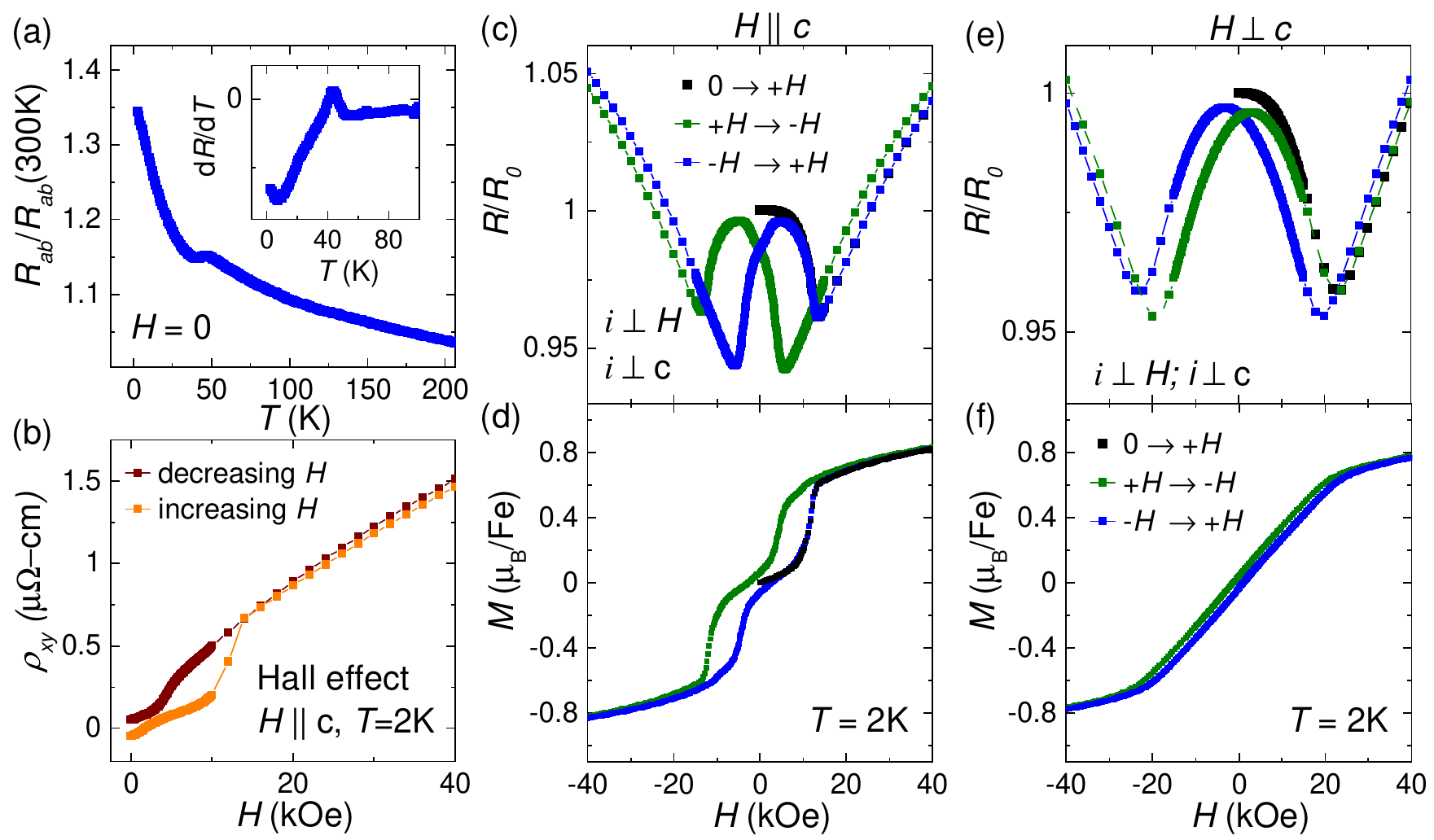}%
\caption{In-plane electrical transport properties of furnace-cooled Fe$_{4.8}$AsTe$_2$ crystals. (a) Relative temperature dependence of the resistivity with inset showing the temperature derivative. (b) Hall resistivity as a function of applied field for increasing and decreasing field (ZFC not shown). (c) Transverse magnetoresistance for field along [001] as a ratio relative to the zero field cooled value of $R_0$ and (d) the corresponding magnetization loop.  Panels (e,f) contain data for a field applied within the basal plane ($H$ $\perp$ $c$) with (e) the transverse magnetoresistance and (f) the magnetization loop.}%
\label{Resist}
\end{figure*}

The electrical transport properties of Fe$_{4.8}$AsTe$_2$ were investigated using in-plane electrical resistivity $\rho$, magnetoresistance (MR) and Hall effect measurements, and the primary results are shown in Fig.\,\ref{Resist}.  The electrical resistivity of our Fe$_{4.8}$AsTe$_2$ crystals increases slightly upon cooling, which differs from the behavior observed in Fe$_{5-x}$GeTe$_2$ ($x$ $\approx$ 0.2) crystals where the resistivity decreases upon cooling.  The room-temperature resistivity of both compounds is fairly similar (hundreds of $\mu \Omega$-cm).   Interestingly, $\rho$ also increases upon cooling in Fe$_{3-x}$AsTe$_2$ whereas Fe$_{3-x}$GeTe$_2$ has bad metal like behavior.\cite{May2016,Verchenko2016new}  As shown in Fig.\,\ref{Resist}a, the resistivity in Fe$_{4.8}$AsTe$_2$ has a small anomaly near the magnetic transition and it increases more rapidly upon cooling below $T_{\rm N}$. This behavior is suggestive of gapping in the Fermi surface caused by the magnetic order.   It would be interesting to probe the extent to which correlations impact the physical properties of Fe$_{5-x}$AsTe$_2$ in comparison to Fe$_{5-x}$GeTe$_2$.

The Hall effect data are shown in Fig.\,\ref{Resist}(b) for $T$=2\,K.  Data are shown for decreasing the field toward zero (red data), which results in a remanent moment and an associated remanent (anomalous) Hall effect.  Data are also shown for increasing the field from this remanent state (orange data), and in the increasing field condition the spin flop appears to have a stronger impact on the observed Hall effect signal.  The anomalous portion is not very large in Fe$_{4.8}$AsTe$_2$, even after the spin-flop transition, which is consistent with the antiferromagnetic order inferred from magnetization measurements.  The ordinary Hall resistance is non-linear with applied field, with curvature decreasing on warming yet still present well-above \TN (see Supporting Materials), which suggests that multiple bands contribute to conduction in Fe$_{4.8}$AsTe$_2$. At 250\,K the Hall resistance is seemingly linear with $H$ (10 to 80\,kOe), and a single carrier analysis yields a Hall carrier density of $\approx$1$\times$10$^{22}$holes/cm$^{3}$. This value, which changes with $T$, is most likely impacted by the existence of multi-carrier transport and thus the Hall data are not a good measure of the metallicity of Fe$_{4.8}$AsTe$_2$ without more detailed knowledge of the Fermi surface.  Holes and electrons both contribute to conduction in Fe$_{4.86}$GeTe$_2$ as well.\cite{May2019}  For comparison sake,  the Hall carrier density calculated by assuming a single band model is $\approx$1.5$\times$10$^{21}$holes/cm$^{3}$ for Fe$_{4.86}$GeTe$_2$ crystals at 375\,K (above the Curie temperature); data taken from Ref. \cite{May2019}.  These results suggest more free holes in Fe$_{4.8}$AsTe$_2$ than in Fe$_{4.86}$GeTe$_2$, though the temperature dependence of the resistivity is less metallic in Fe$_{4.86}$GeTe$_2$ and this may point to possible scattering effects.  Of course, these numbers may be skewed by electron-hole compensation effects in the Hall effect that artificially raise the single-band carrier density.

The magnetoresistance (MR) data were collected in two transverse configurations with current always flowing within the $ab$-plane and always perpendicular to the applied field, which is either directed along [001] or orthogonal to [001]; additional schematic illustrations are provided in the Supporting Materials.  The data for $H$ $\parallel$ $c$ are shown in Fig.\ref{Resist}(c) and data for $H$ within the basal plane (yet still perpendicular to the current) are shown in Fig. \ref{Resist}(e).  The corresponding magnetization loops are presented in Figs.\,\ref{Resist}(d,f) to illustrate how the magnetic hysteresis is coupled to the electrical resistivity.  Starting from a zero field cooled state ($R_0$), the application of a magnetic field decreases the resistivity of Fe$_{4.8}$AsTe$_2$ and thus negative magnetoresistance is observed.  Above a critical field, which varies with orientation, the sign of d$R$/d$H$ changes and positive MR is observed at large fields.  Together with the presence of strong magnetic hysteresis, this leads to butterfly-shaped magnetoresistance loops. These loops also reveal that the remanent moment leads to negative magnetoresistance at $H$=0 relative to the ZFC value of $R_0$. Upon further demagnetizing the sample and passing beyond the coercive field, the resistance approaches the $R_0$ value before the negative MR takes over and causes a local minimum near the critical field.  The net result is the butterfly-shaped resistance loop.  This behavior is most dominant for $H$ $\parallel$ $c$ with the spin-flop, but can also be observed for $H$ $\perp$ $c$ where the remanent moment and hysteresis are much smaller.  Interestingly, butteryfly-shaped hysteresis loops have recently been reported for ultra-thin Fe$_{5-x}$GeTe$_2$ where a thickness effect appears to be important.\cite{Ohta2021butterfly}

The negative MR observed for small applied field is likely caused by alignment of   moments within a magnetic domain or alignment of the magnetic domains.  The trend towards positive MR starts at the spin flop ($\approx$10kOe) and MR ultimately reaches 10\% for $H$ $\parallel$ $c$ at 90\,kOe and 2\,K and slightly lower for $H$ $\perp$ $c$.  Positive MR is typical of nonmagnetic or paramagnetic metals and is also observed in some antiferromagnetic materials, but is not typical of a ferromagnet where the applied field typically suppresses fluctuations and aligns domains to reduce carrier scattering (especially near $T_{\rm C}$).  Positive MR can also be observed above the saturation field in a ferromagnet, as in Fe$_{5-x}$GeTe$_2$ at low $T$ and high field.\cite{May2019}  In Fe$_{4.8}$AsTe$_2$, the likely existence of multiple bands at the Fermi level complicates the interpretation of the magnetoresistance, though the anisotropic behavior and butterfly-shaped hysteresis loops demonstrate a strong coupling of the magnetism to the electronic transport.

\section{Conclusions}

The impact of arsenic substitution for Ge in the high-Curie temperature vdW material Fe$_{5-x}$GeTe$_2$ was probed and the properties of Fe$_{4.8}$AsTe$_2$ were reported.  Small additions of As appear to enhance the ferromagnetism in polycrsytalline Fe$_{5-x}$GeTe$_2$ by suppressing spin fluctuations on the Fe1 sublattice. This also decreases the anisotropy field and thus provides a means for local tuning of the magnetism without major lattice changes. However, large concentrations of As lead to a significant decrease in the Curie temperature and saturation moment. While structural characterization of Fe$_{4.8}$AsTe$_2$ by x-ray diffraction was hindered due to the presence of stacking disorder and potential phase separation, a key similarity to Fe$_{5-x}$GeTe$_2$ is evidenced by observation of an in-plane supercell associated with occupancy/vacancy order on the Fe1 sublattice.  Electron diffraction and microscopy investigations are necessary to inspect the local crystallography and phase evolution in this complex material.  Crystals of Fe$_{4.8}$AsTe$_2$ that are slowly cooled display characteristics of long-range antiferromagnetic order with a small ferromagnetic component, while quenched crystals display characteristics of glassy magnetism.  In the magnetically ordered phase, a spin-flop transition demonstrates the dominant easy-axis [001] anisotropy of the moments, and this meta-magnetic transition is strongly coupled to the electrical transport properties and causes butterfly-shaped magnetoresistance loops.  In total, these results motivate detailed experimental and theoretical efforts to identify dopants that lead to enhanced magnetic ordering temperatures and anisotropy control in itinerant vdW magnetic materials.  Importantly, this work demonstrates that small concentrations of such dopants need to be considered due to the sensitivity of itinerant magnetic materials to small changes in the Fermi energy or crystal chemistry.

\section{Methods}

Single crystals of Fe$_{4.8}$AsTe$_2$ were grown by heating the pure elements (Fe, As, Te, I) in an evacuated silica ampoule to 750\degrees C over 30\,h followed by a dwell period of approximately 10\,d. The largest crystals were 2-3\,mm in lateral dimension, and these were obtained from growths performed with a hot-side temperature of 750\degrees C in a horizontal tube furnace.  The growth ampoules were either allowed to cool in the furnace over 8-12\,h or were quenched into an ice-water bath.  For quenched crystals, iodine was rinsed from the crystals using alcohols and/or acetone to prevent accelerated tarnishing due to the hygroscopic nature of iodine; during slow cooling the iodine deposits on the silica ampoule due to the temperature gradient and rinsing is not required.

Polycrystalline samples for the solid-solution series Fe$_{5}$Ge$_{1-y}$As$_y$Te$_2$ (with nominal $y$= 0.025, 0.05, 0.25, 0.50, 0.75) were synthesized by first reacting the elements at 750\degrees C for 72\,h using an initial heating rate of 25\degrees/h.  The reacted products were ground briefly in air, pressed into pellets with a diameter of one-half inch, and then sealed in silica ampoules with a small pressure of argon.  A second heat treatment at 750\degrees C lasted for approximately 200\,h prior to quenching into an ice-water bath.  Polycrystalline samples of nominal compositions  Fe$_{4.8}$AsTe$_2$, Fe$_{4.5}$AsTe$_2$ and Fe$_{5.5}$AsTe$_2$ were synthesized in a similar manner and the impact of thermal processing (quenching, cooling in furnace) was examined for these samples.  Le Bail fitting was performed in FullProf\cite{FullProf} to obtain lattice parameters, though it is noted that the fits are of low quality due to asymmetric and inconsistent broadness in diffraction intensities associated primarily with stacking faults, though the data may also be impacted by  phase separation issues.

Chemical analysis of slow-cooled, vapor transport grown Fe$_{4.8}$AsTe$_2$ crystals was performed using wavelength dispersive spectroscopy (WDS) in a JEOL 8200 with elemental standards of Fe, Te and binary InAs.  The beam energy was 25\,keV using a current of 50\,nA.

Single-crystal x-ray diffraction data were collected at 220\,K using a Bruker D8 Quest with a nitrogen cold stream while the crystals were mounted on a kapton loop using paratone oil.  Structural modeling was performed using ShelX after data reduction via Bruker's APEX3 software.\cite{Sheldrick2015}  Crystals ($<$70$\mu$m) were selected from the products of growths that started with different nominal compositions (Fe$_5$AsTe$_2$ and Fe$_6$AsTe$_2$) and different thermal histories (quenching, furnace cooling).  Quenched crystals were found to have less streaking along $l$ in the relevant reciprocal space maps of the diffracted intensity, suggesting they have fewer stacking faults than the furnace cooled crystals.  Data from the heavily faulted crystals could not be refined, and refinement results from the structural solution for the quenched crystals are provided in the Supporting Materials. X-ray diffraction data were collected using a PANalytical X'Pert Pro MPD with a Cu K$\alpha_1$ ($\lambda$=1.5406\,\AA) incident beam monochromator.  Some degradation of the diffraction data was observed after several hours of exposure, and thus these samples are mildly sensitive to moisture and/or oxygen.

Transport measurements were performed in a Quantum Design Dynacool.  The Hall effect data were anti-symmetrized (odd only) to avoid mixing of the transverse and longitudinal signals due to imperfect measurement geometry.  Transport data (MR, Hall) for $H \parallel c$ were collected simultaneously using a six-wire method while data for $H \perp c$ were collected on a separate crystal.  Magnetization measurements were collected in SQUID magnetometers (MPMS-XL and MPMS3) from Quantum Design (QD) and ac susceptibility data were collected in a QD PPMS and the MPMS3.  The contour plots of d$M$/d$H$ shown in Fig.\,\ref{MHparC}(b,c) were obtained using temperature steps of 1\,K and the applied field was stabilized in steps of 100\,Oe.  The data in Fig.\,\ref{MHparC}(b) were collected upon increasing $H$ after cooling from 150\,K in zero field.  The data in Fig.\,\ref{MHparC}(c) were obtained while decreasing $H$ from 70\,kOe with data first collected at 2\,K, then 3\,K, and so on.

Density functional theory (DFT) calculations were performed using the Perdew-Burke-Ernzerhof  (PBE) exchange-correlation functional\cite{PerdewGGA1996} as implemented in the VASP code.\cite{Kresse1996a} The kinetic energy cutoff of the plane-wave basis is 268\,eV; changing to a cutoff energy of 400\,eV resulted in quantitative changes of less than 10\% and no qualitative changes. The projector augmented wave method is used to describe the interaction between ions and electrons.\cite{Kresse1999}  A 6 $\times$ 6 $\times$ 1 k-point mesh was used for a 2 $\times$ 2 $\times$ 2 supercell. The crystallographic parameters are fixed at the experimentally measured values for quenched crystals but the atomic positions are optimized until the force on each atom is less than 0.01 eV/\AA~.  It is again emphasized that these calculations are highly idealized and point and planar defects may be essential in understanding the different magnetic phases.

\section{Acknowledgments}
We thank R. Custelcean for assistance with x-ray single-crystal diffraction measurements and M. Lance for assistance with WDS measurements. This work was supported by the U. S. Department of Energy, Office of Science, Basic Energy Sciences, Materials Sciences and Engineering Division.

\providecommand{\newblock}{}

\pagebreak
\newpage
\pagebreak
\begin{center}
\textbf{\large Supporting Materials: Tuning the Room Temperature Ferromagnetism in Fe$_{5}$GeTe$_2$ by Arsenic Substitution}
\end{center}
\setcounter{equation}{0}
\setcounter{figure}{0}
\setcounter{table}{0}
\setcounter{page}{1}
\makeatletter
\renewcommand{\theequation}{S\arabic{equation}}
\renewcommand{\thefigure}{S\arabic{figure}}

\begin{abstract}
This Supporting Material document provides additional data regarding the characterization of crystal structure and physical properties in polycrystalline samples of nominal composition Fe$_5$Ge$_{1-x}$As$_x$Te$_2$ and single crystals of Fe$_{5-x}$AsTe$_2$ ($x$ $\approx$ 0.2).  Single crystal and powder x-ray diffraction data are presented along with magnetization and specific heat data.  The impact of synthesis conditions on the observed phases and magnetic properties of Fe$_{5-x}$AsTe$_2$ samples is also shown; thermal quenching leads to a state with glassy character while cooling in the furnace leads to behavior consistent with canted antiferromagnetic order.  Stacking disorder and phase competition is present in all samples, but is less evident in the diffraction data from thermally quenched samples.
\end{abstract}

\pagebreak

\newpage

\setcounter{equation}{0}
\setcounter{figure}{0}
\setcounter{table}{0}
\setcounter{page}{1}
\makeatletter
\renewcommand{\theequation}{S\arabic{equation}}
\renewcommand{\thefigure}{S\arabic{figure}}


\textbf{Density Functional Theory Calculations}\\ 

Spin-polarized density functional theory (DFT) calculations were used to inspect the idealized magnetic ground state in Fe$_5$AsTe$_2$ with an emphasis on comparing to calculations previously reported for Fe$_5$GeTe$_2$.[1]  The structural data in Table \ref{tab:refine2} were utilized.  These structural data (lattice parameters, atomic positions) were obtained from single crystal x-ray diffraction data collected on a quenched single crystal; strong evidence for long-range magnetic order was not obtained for quenched \FASx crystals but AFM correlations are inferred by the magnetization data.\\

First, the impact of different occupancy patterns for the Fe1 sublattice were inspected using ferromagnetic moment alignment as discussed previously for Fe$_5$GeTe$_2$, (occupancy patterns leading to so-called checkerboard Fe1a,b; zig-zag Fe1a,b; and all-up with only Fe1a occupied).[1] Appropriate supercells were considered. Similar to previous calculations for Fe$_5$GeTe$_2$,[1] the results suggest that atomic configurations where Fe occupies the Fe1a and Fe1b sites are close in energy whereas occupancy of only Fe1a positions is not energetically favorable.  Thus, a distribution of Fe across the Fe1a,b positions is expected and this will lead to local disorder and short-range order in the real crystals. After imposing the checkerboard pattern for occupancy on the Fe1a,b sublattice, a ferromagnetic ground state was found to have the lowest energy.  In this state, the different sublattice positions carry different moments and because of the reduction in symmetry for (the checkerboard Fe1a,b) supercell the `same' sites can carry different moments. These different moments are shown in the main text in Table 1. In Fe$_5$AsTe$_2$, the moments are found to vary somewhat significantly on the Fe2 sublattice depending on whether or not they are close (large) or far (smaller moments) from the occupied Fe1a,b site.  This demonstrates the importance of local chemistry effects that may have a strong impact on the macroscopic magnetic properties.\\

\providecommand{\newblock}{}


\pagebreak 



\begin{figure*}[ht!]%
\includegraphics[width=1.95\columnwidth]{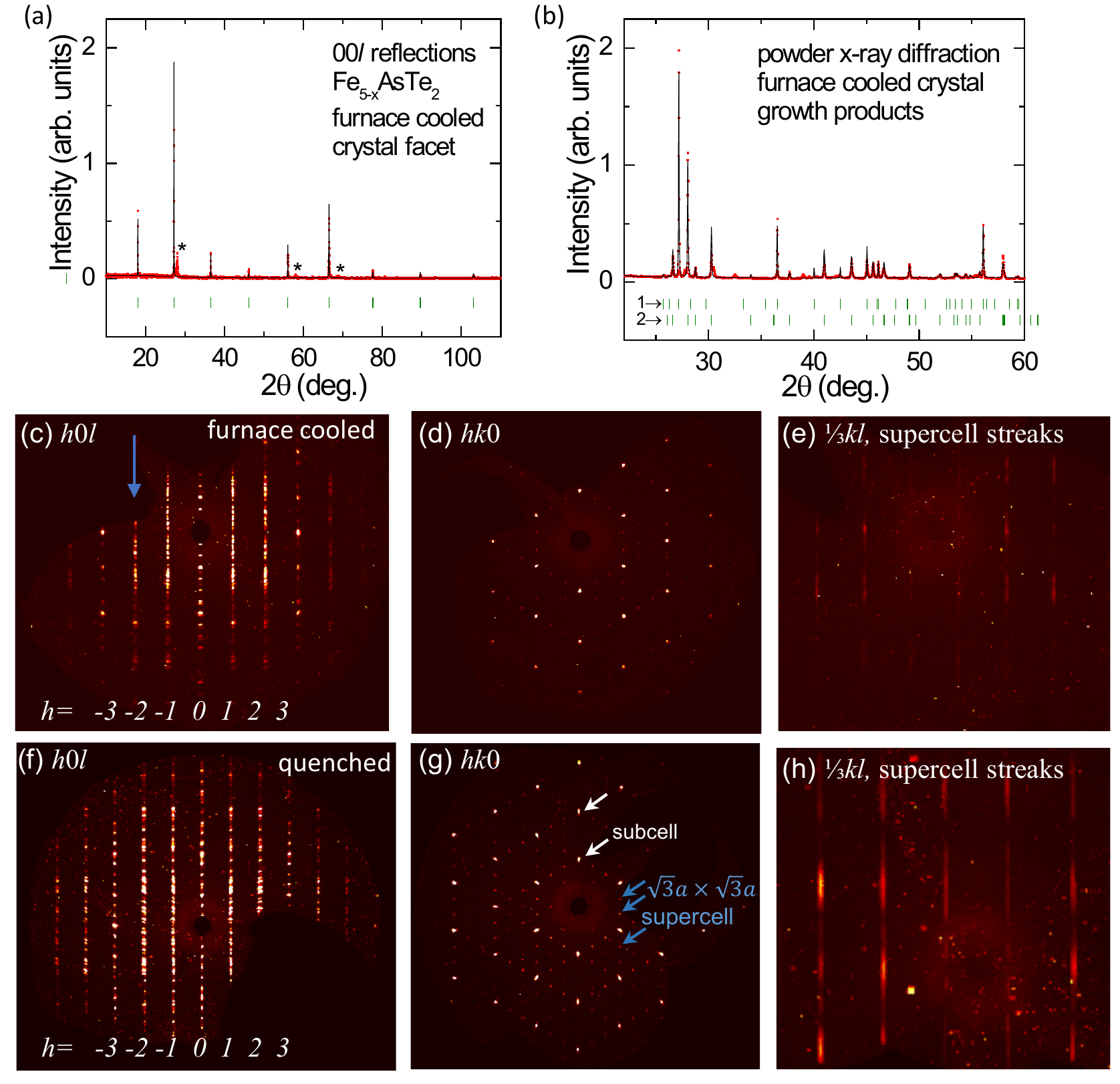}%
\caption{(a) Diffraction from Fe$_{4.8}$AsTe$_2$ crystal facet (b) powder x-ray diffraction from products of crystal growth with data fitted by two phases of the same rhombohedral symmetry, with the primary phase displaying characteristics of a large amount of stacking disorder and larger lattice parameters.  (c,d,e) Precession images from single crystal x-ray diffraction data for a furnace-cooled Fe$_{4.8}$AsTe$_2$  crystal. (f,g,h) Precession images from single crystal x-ray diffraction data for a quenched crystal.  Cooling over 8-12\,h in the furnace results in more streaking along $l$ for $h0l$ reflections with $h$ $\neq$ 0, such as at the position of the blue arrow in panel (c), and this is associated with stacking disorder.  Such disorder makes it difficult to index and integrate the data to obtain a structural solution.  Less streaking is observed in (f) for the quenched crystal.  Both types of crystals displayed evidence for an in-plane supercell (d,g).  The data in (e,h) demonstrate that the superlattice peaks are smeared along $l$ which shows that the superlattice is very incoherent along $l$. The short range order is expected to be an in-plane effect associated with occupancy of the Fe1a,b sites.}%
\label{XRD48}
\end{figure*}

\pagebreak

\begin{table*}[ht!]
\caption{\textbf{Refined structural parameters for \FAS from single-crystal x-ray diffraction data collected on a \textit{quenched crystal}.} Space group $R\bar{3}m$ (No. 166); $a$ = 4.0088(6)\AA, $c$ = 29.279(6)\AA, $T$ = 220\,K, $R1$  =  0.0630, $wR2$ = 0.150 and GooF = 1.34 for all 325 reflections after merging using 17 parameters. As1 and Fe1 are splits sites with occupancy limited to 50\%; the displacement parameters for Fe1 and Fe2 were constrained to be equal. All atomic positions are at 0,0,$z$ with $z$ provided (6c Wyckoff position).}
\begin{tabular}[c]{|c|c|c|c|c|}
\hline
atom  &  $z$ &  occupancy &  & \\
\hline
Te1 &  0.21875(4)  &  1 &      &\\
As1 &  0.01033(13) &  0.5 & (split site)&\\ 
Fe1 &  0.0700(2)   &  0.402(11) & (split site) &\\ 
Fe2 &  0.30965(9)  &  1 &  &\\
Fe3 &  0.39892(11) &  1 & &\\ 
\hline
 &  $U_{11}$ &$U_{22}$ & $U_{33}$ & $U_{12}$  \\
\hline
Te1 & 0.0125(4)  & 0.0125(4)  & 0.0150(6)   & 0.0062(2)  \\
As1 & 0.0055(8)  & 0.0055(8 ) & 0.026(2)    & 0.0028(4\\
Fe1 & 0.0139(7)  & 0.0139(7 ) & 0.0139(10)  & 0.0070(3)\\  
Fe2 & 0.0139(7)  & 0.0139(7)  & 0.0139(10)  & 0.0070(3) \\
Fe3 & 0.0307(11) &0.0307(11)  & 0.0164(12)  & 0.0154(6) \\
\hline
\end{tabular}
\label{tab:refine2}
\end{table*}

\pagebreak

\begin{figure*}[ht!]%
\includegraphics[width=\columnwidth]{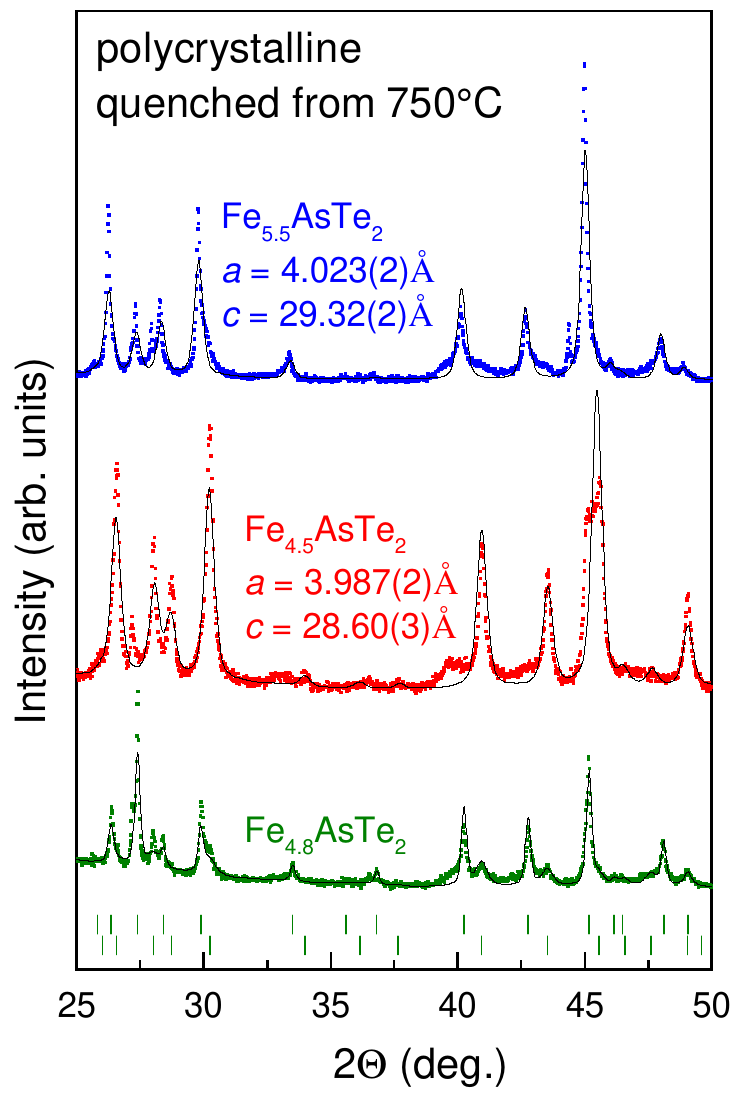}%
\includegraphics[width=\columnwidth]{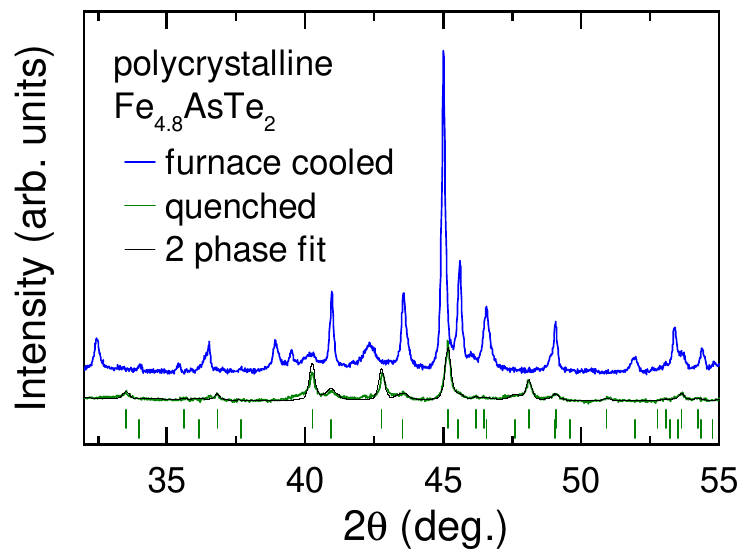}%
\caption{(left) Powder diffraction data collected at ambient condition and Le Bail fits (black lines) for Fe$_{5}$AsTe$_2$ type samples at different \textit{nominal} compositions as indicated in the figure.  For a starting composition of Fe$_{4.8}$AsTe$_2$, two phases are observed and the lattice parameters are similar to those observed in the reactions performed at compositions Fe$_{4.5}$AsTe$_2$ and Fe$_{5.5}$AsTe$_2$.  (right) Impact of quenching or slow cooling on the diffraction data of a polycrystalline sample at a nominal composition of Fe$_{4.8}$AsTe$_2$.  In the furnace cooled sample, the dominant phase with larger lattice parameters ($a$=4.0234(14)\AA\,, $c$=29.49(1)\AA\,) displays more significant peak broadening associated with stacking faults than does the phase with smaller lattice parameters ($a$=3.9740(16)\AA\,, $c$=28.63(2)\AA\,) and the $c$/$a$ ratio is larger for the larger unit cell phase.}%
\label{XRD2}
\end{figure*}

\begin{figure*}[h!]%
\includegraphics[width=1.95\columnwidth]{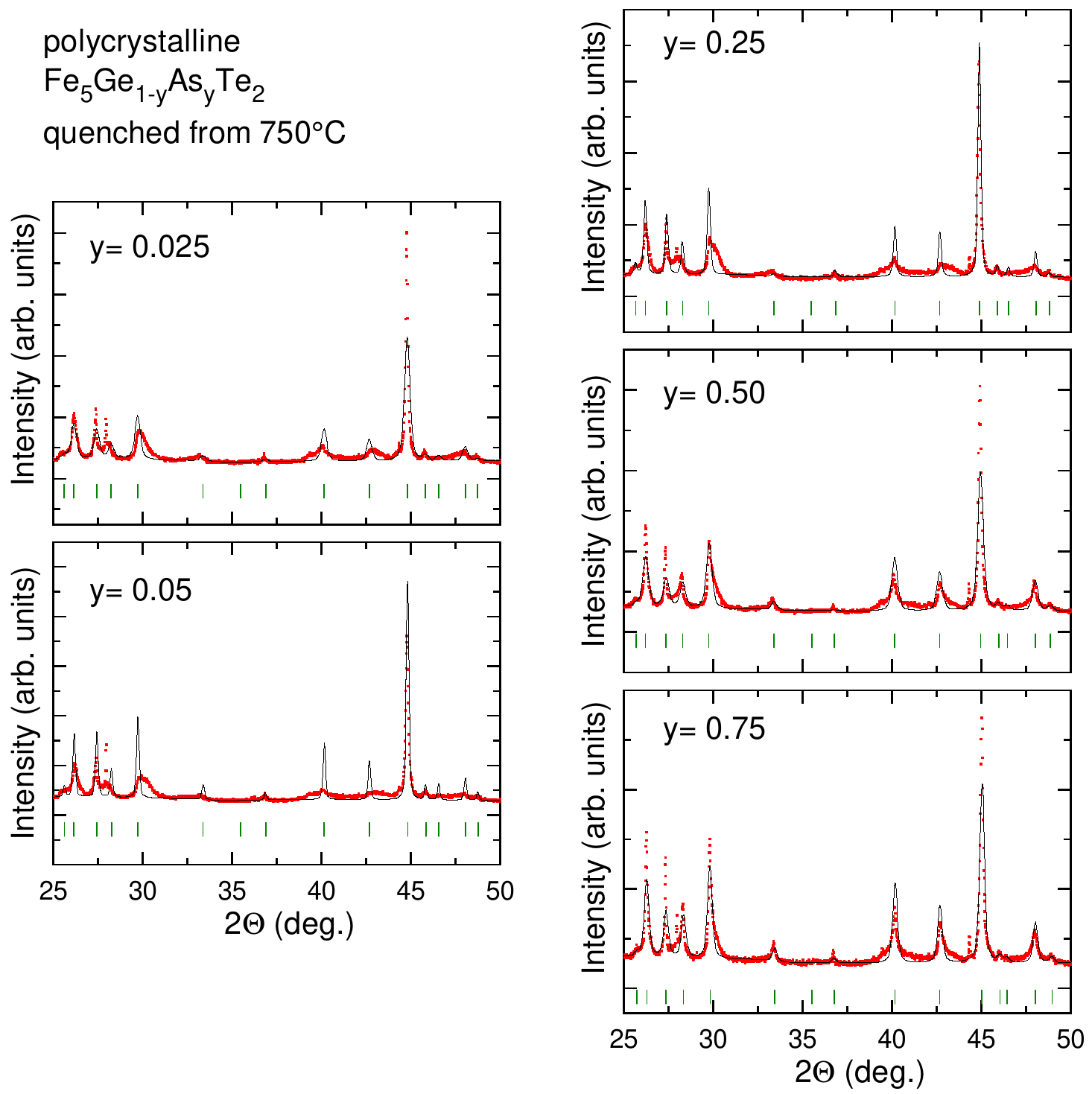}%
\caption{X-ray powder diffraction data for polycrystalline samples of nominal composition Fe$_{5}$Ge$_{1-y}$As$_y$Te$_2$ obtained after quenching from 750\degrees C. Peak broadening caused by stacking faults is the strongest in samples with the lowest arsenic content $y$ for the mixed Ge-As powder samples. For instance, the peak near 40\degrees\, 2$\theta$ is the 1\,0\,10 Bragg peak (impacted by stacking faults) and the strongest peak (near 45\degrees) is the 1\,1\,0 Bragg peak (not impacted by stacking faults).}%
\label{PXRD_alloy}
\end{figure*}

\pagebreak
\newpage



\pagebreak

\begin{figure}[h!]%
\includegraphics[width=0.95\columnwidth]{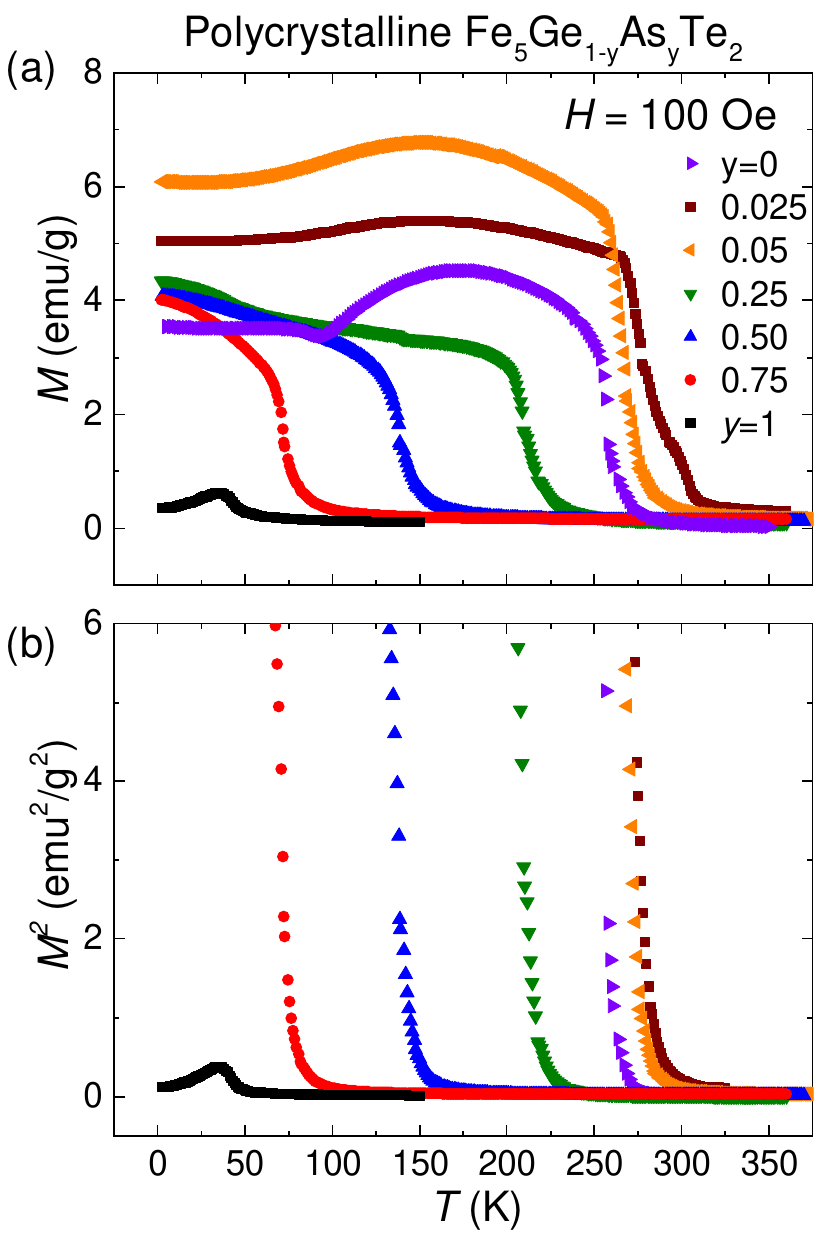}%
\caption{(a) Low-field magnetization data for polycrystalline Fe$_{5}$Ge$_{1-y}$As$_y$Te$_2$ samples utilized to obtain the magnetic ordering temperatures. For $0 \leq y \leq 0.75$ with dominant ferromagnetic behavior the Curie temperatures were obtained by taking an average of two methods: (1) the junction of linear extrapolations from above and below the apparent $T_C$ using the squared magnetization $M^2$ versus $T$ and (2) from the peak in the derivative of d$M$/d$T$. The $M^2$ versus $T$ data are shown in panel (b).  The difference in these approaches was 2 - 5\,K and the qualitative results presented are independent of the approach.  The $y$=0.025 sample displays a small onset in magnetization at slightly higher $T$ than the reported $T_C$, perhaps due to inhomogeneity.}%
\label{CurieTempData}
\end{figure}

\pagebreak

\begin{figure*}[h!]%
\includegraphics[width=0.95\columnwidth]{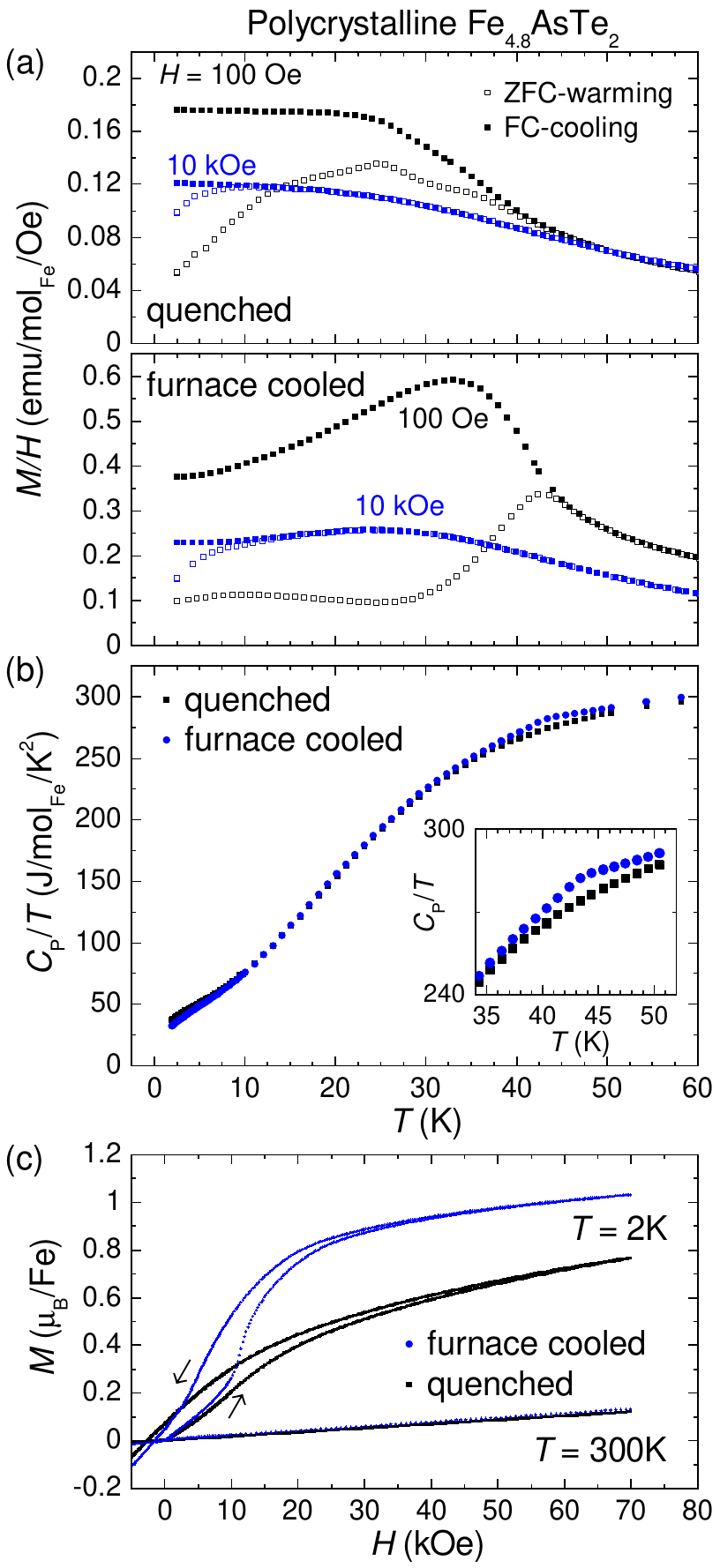}%
\caption{Impact of synthesis conditions (thermal quenching, furnace cooling) on the magnetic properties of polycrystalline samples of nominal composition Fe$_{4.8}$AsTe$_2$.  (a) Temperature-dependent magnetization for the quenched (upper panel) and furnace cooled (lower panel) samples with zero field cooled (ZFC) and field cooled (FC) conditions shown.  (b) Specific heat data near the magnetic transition with inset having the same $y$-axis units as the main panel. (c) Isothermal magnetization data, with increasing and decreasing magnetic field shown for $T$=2\,K. The shape of $M$($T$) data in the upper panel (a) suggest the magnetism in the thermally quenched crystals is likely glassy at low $T$, with $M$ becoming essentially independent of $T$ below the apparent freezing temperature.  By contrast, $M$($T$) in the furnace-cooled crystals has a maximum and decreases upon cooling into an antiferromagnetic state, consistent with the magnetic anisotropy and spin flop transition that is also observed (see main text).  The polycrystalline samples were obtained by first reacting the elements at 850\degrees C for 228h and this reaction was quenched in an ice-water bath.  The products were ground in air, pressed into a pellet, and the pellet was divided for further annealing.  The different pieces were again sealed in SiO$_2$ ampoules and then annealed at 750\degrees C for 160h before being either quenched or cooled in the furnace.  The samples were then utilized for these measurements and the corresponding diffraction data are presented in Fig. S2(right).}%
\label{SynthesisImpact}
\end{figure*}

\pagebreak

 \begin{figure}[h!]%
\includegraphics[width=0.95\columnwidth]{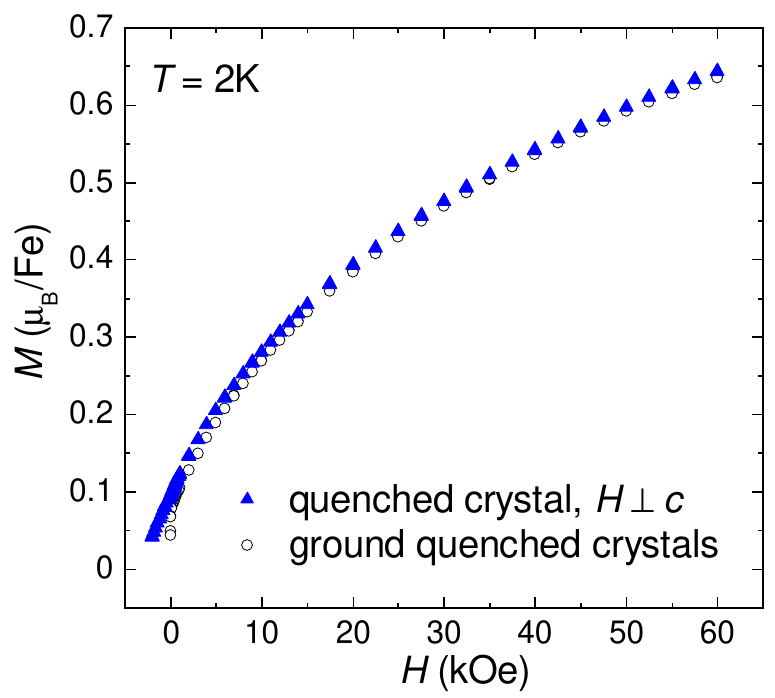}%
\caption{Isothermal magnetization data for thermally quenched crystals of Fe$_{4.8}$AsTe$_2$ demonstrating the lack of magnetic anisotropy by comparing the data for an oriented single crystal to data for ground crystals.  Together with the observed $M$($T$) behavior, the magnetic isotropy is generally consistent with a lack of long-range magnetic order, possibly due to a glassy ground state in the thermally quenched crystals.}%
\label{MHquenched}
\end{figure}

\pagebreak

\begin{figure*}[h!]%
\includegraphics[width=1.95\columnwidth]{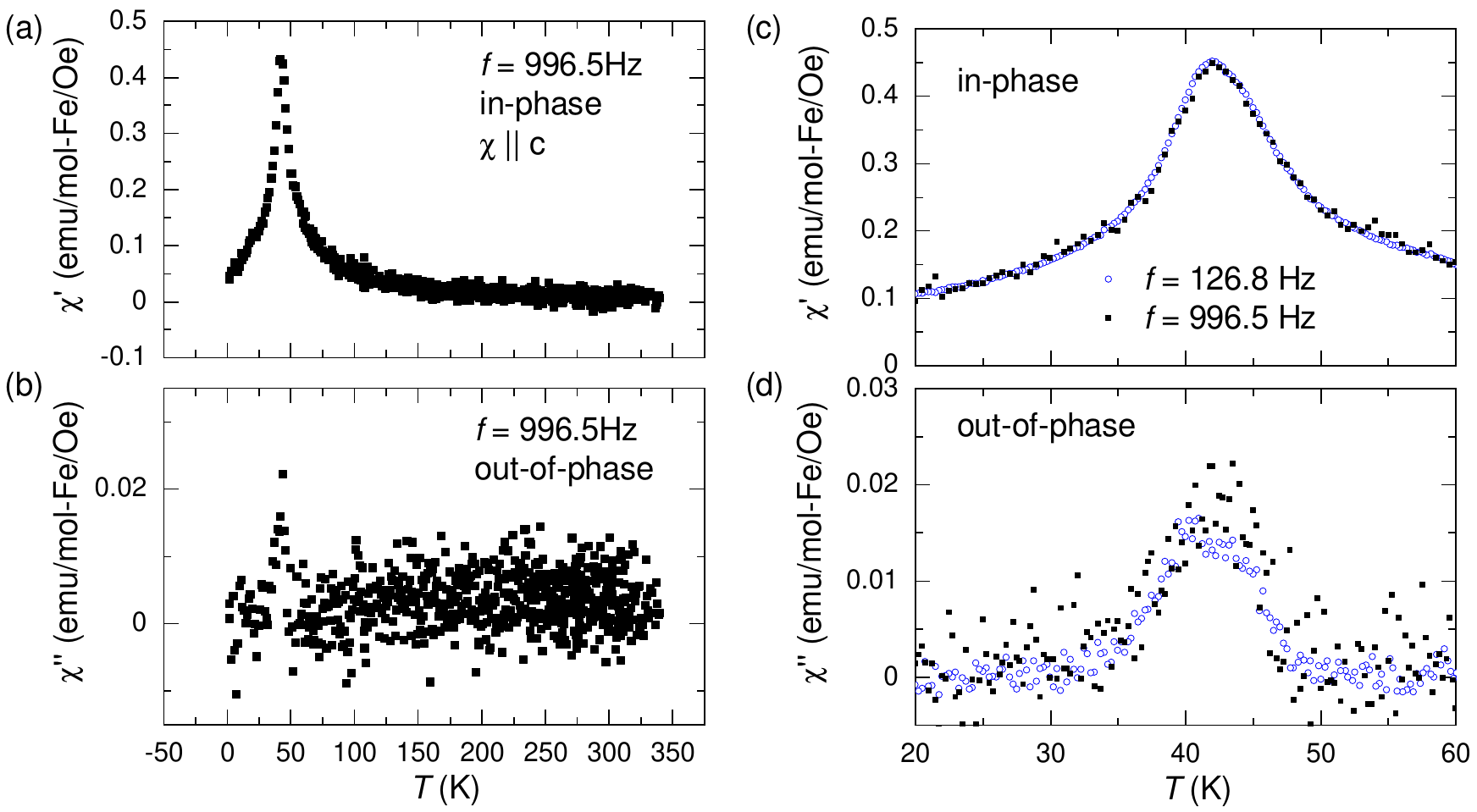}%
\caption{AC magnetic susceptibility data collected upon cooling with zero applied DC magnetic field for a furnace-cooled Fe$_{4.8}$AsTe$_2$ crystal. An AC drive of 2\,Oe along the $c$-axis was utilized. A single temperature-dependent transition is observed at \TN $\approx$ 42\,K and the temperature of this maximum in $\chi'$ does not notably shift with frequency $f$ for the frequencies inspected.  Data with finer temperature spacing were collected for different $f$ near \TN and are shown in (c,d).  The out-of-phase contribution (b,d) is very small relative to the in-phase contribution.  The existence of a small out-of-phase contribution is consistent with the non-compensated nature of the antiferromagnetic order that is inferred from DC measurements.}%
\label{ACHparC}
\end{figure*}

\pagebreak 

\begin{figure}[h!]%
\includegraphics[width=0.95\columnwidth]{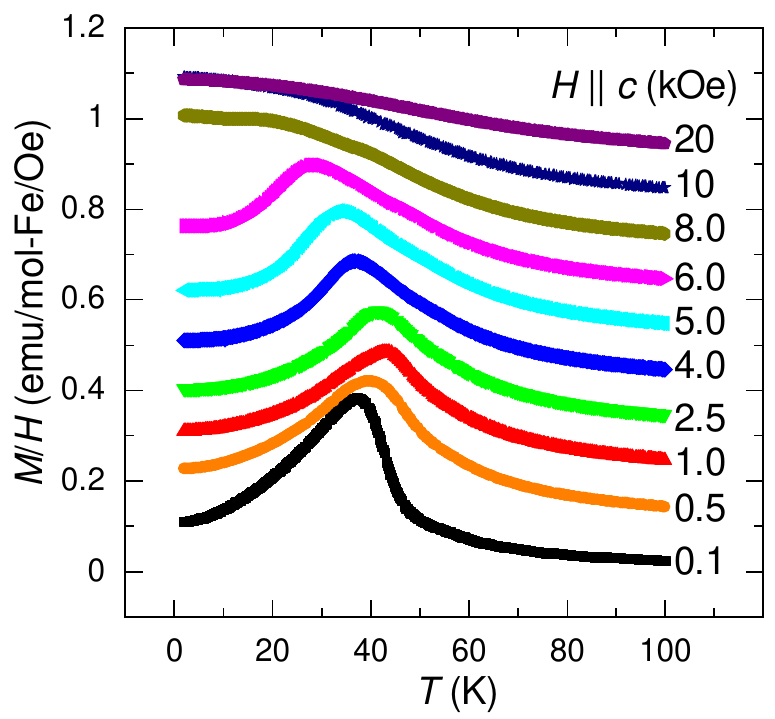}%
\caption{Magnetization data for a furnace-cooled Fe$_{4.8}$AsTe$_2$ crystal: $M$ divided by applied field $H$ upon cooling in different applied fields as listed.  The data for $H$=0.1\,kOe were scaled by 0.5 and the remaining data sets were continually shifted along the y-axis by 0.1 for clarity.  These data demonstrate the shifting of the cusp in $M$($T$) to higher temperatures for increasing applied field up to approximately 4\,kOe when the field is applied along the $c$-axis.  This behavior is consistent with a canted AFM phase.}%
\label{MTfields}
\end{figure}

\pagebreak

\begin{figure*}[h!]%
\includegraphics[width=0.95\columnwidth]{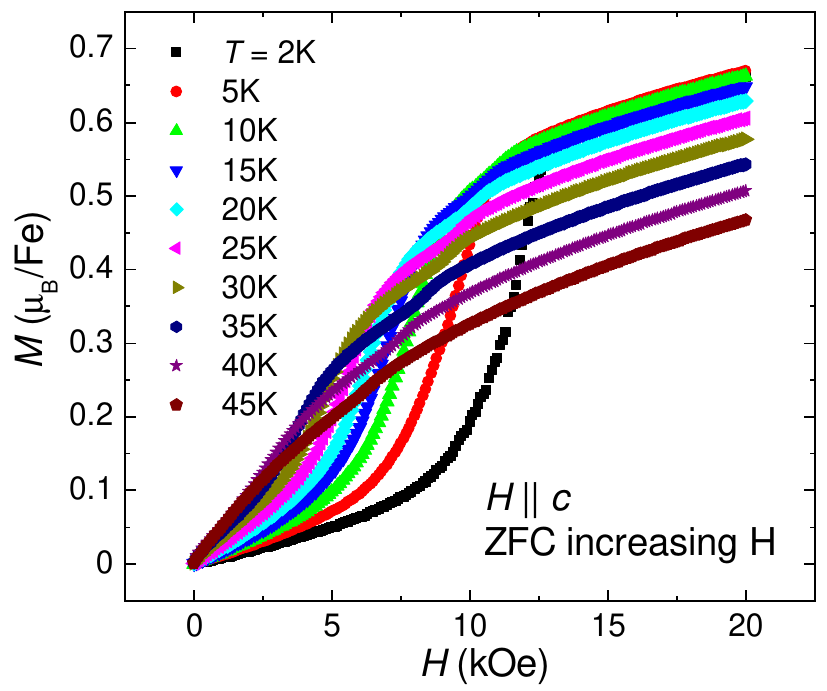}%
\includegraphics[width=0.95\columnwidth]{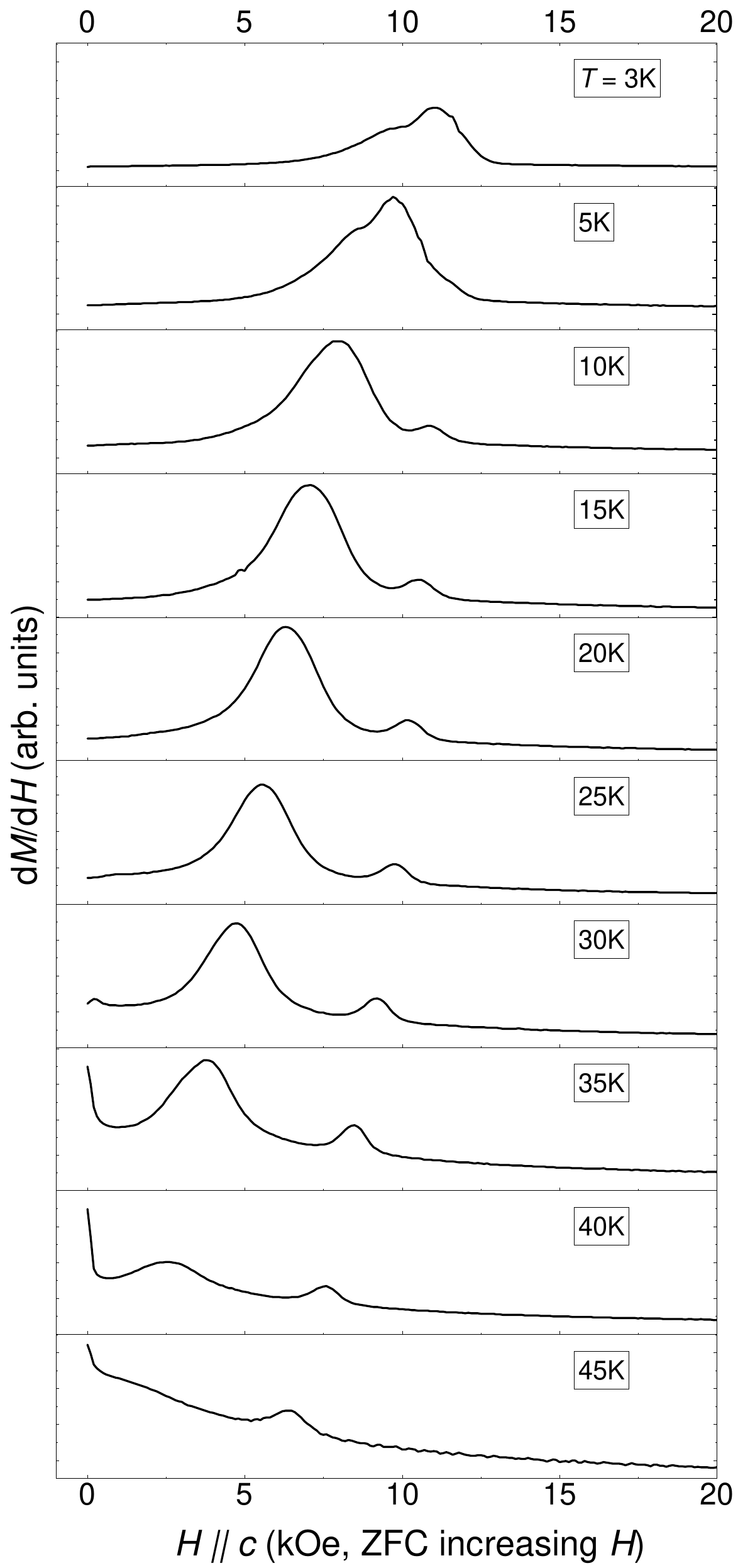}%
\caption{Select isothermal magnetization data for furnace-cooled  Fe$_{4.8}$AsTe$_2$ crystals at different temperatures after zero field cooling (ZFC). (left) Isothermal magnetization and (right) corresponding d$M$/d$H$ data.  Such data were utilized to generate the contour plots in the main text (Fig. 4(b,c)).}%
\label{MHdata}
\end{figure*}

\pagebreak

\begin{figure*}[ht!]%
\includegraphics[width=1.95\columnwidth]{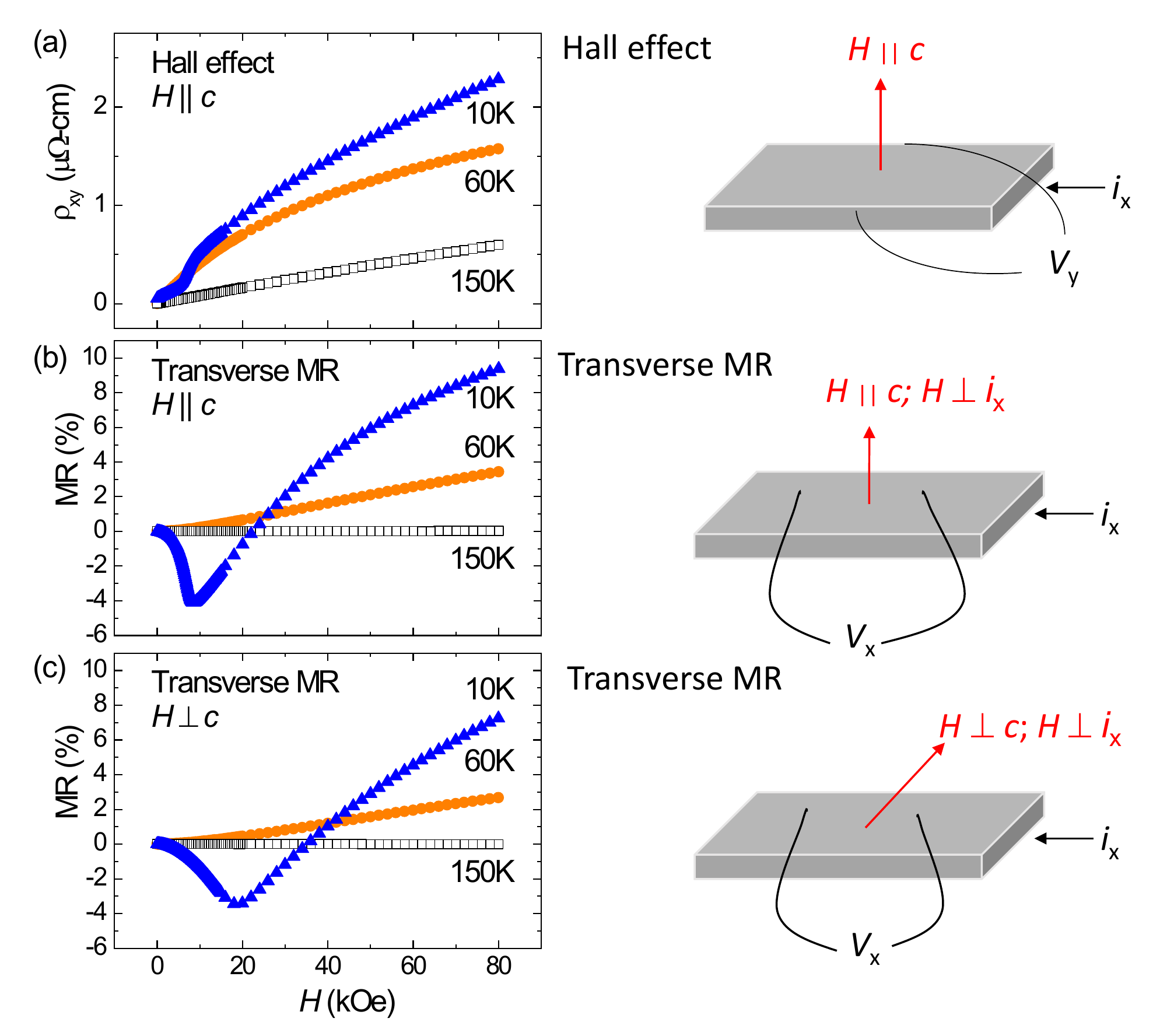}%
\caption{Field-dependent transport data at select temperatures above and below the N\'{e}el temperature of $\approx$42\,K for furnace-cooled crystals of Fe$_{4.8}$AsTe$_2$.  (a) Hall resistivity, (b,c) transverse magnetoresistance MR = $\frac{\rho(H) - \rho(H=0)}{\rho(H=0)}$; here, $\rho$($H$=0) is obtained after demagnetizing the sample to a state with a small remanent moment. The schematics to the right of each panel illustrate the relative orientations of the applied field $H$, current $i$ and measured voltage $V$.  The data in (a) were anti-symmetrized (odd only) while the data in (b,c) were symmetrized (even only) to avoid mixing of transverse and longitudinal voltage responses.  The data were obtained simultaneously, with data in (a,b) obtained on one crystal using six wires while the data in (c) were collected using a different crystal.}%
\label{Transport}
\end{figure*}

\pagebreak

\begin{figure}[ht!]%
\includegraphics[width=0.95\columnwidth]{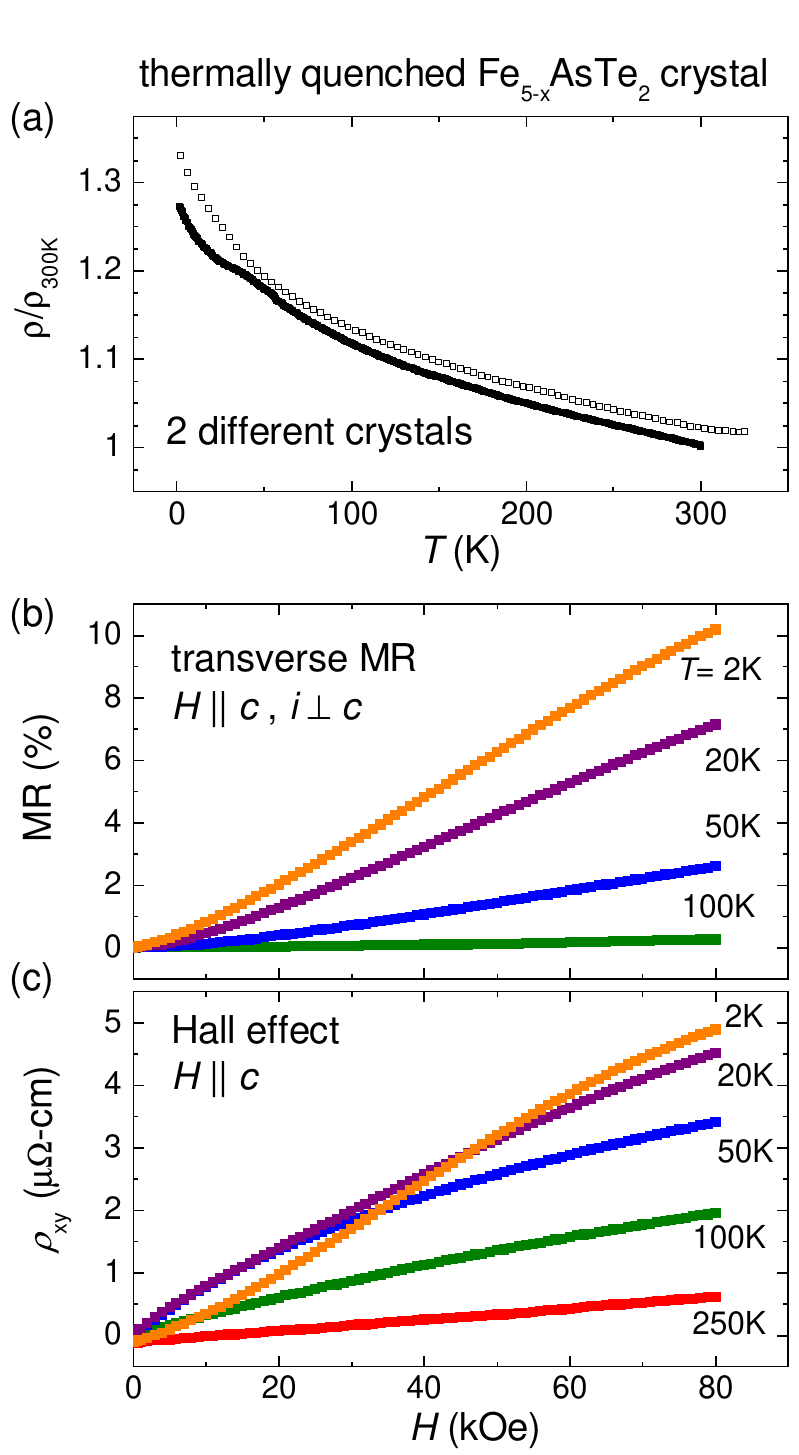}%
\caption{Electrical transport data for thermally-quenched Fe$_{4.8}$AsTe$_2$ crystals.  The butterfly-shaped magnetoresistance associated with the spin-flop transition in the furnace-cooled crystals is not observed in the quenched crystals, which only display positive magnetoresistance at low temperatures.}%
\label{TransportQuenched}
\end{figure}

\pagebreak

%

\begin{thebibliography}{10}

\bibitem{Burch2018}
Burch K~S, Mandrus D and Park J~G 2018 {\em Nature\/} {\bf 563} 47
  

\bibitem{Wang2020rev}
Wang M~C, Huang C~C, Cheung C~H, Chen C~Y, Tan S~G, Huang T~W, Zhao Y, Zhao Y,
  Wu G, Feng Y~P {\em et~al.\/} 2020 {\em Annalen der Physik\/} {\bf 532}
  1900452

\bibitem{Huang2020rev}
Huang P, Zhang P, Xu S, Wang H, Zhang X and Zhang H 2020 {\em Nanoscale\/} {\bf
  12} 2309--2327

\bibitem{Deiseroth2006}
Deiseroth H~J, Aleksandrov K, Reiner C, Kienle L and Kremer R 2006 {\em Eur. J.
  Inorg. Chem.\/} {\bf 2006} 1561

\bibitem{Stahl2018}
Stahl J, Shlaen E and Johrendt D 2018 {\em Z. Anorg. Allg. Chem.\/} {\bf 644}
  1923
  

\bibitem{May2019acs}
May A~F, Ovchinnikov D, Zheng Q, Hermann R, Calder S, Huang B, Fei Z, Liu Y, Xu
  X and McGuire M~A 2019 {\em ACS Nano\/} {\bf 13}(4) 4436
  

\bibitem{Jothi2019}
Jothi P, Scheifers J, Zhang Y, Alghamdi M, Stekovic D, Itkis M, Shi J and Fokwa
  B 2020 {\em Phys. Status Solidi RRL\/} {\bf 14} 1900666
 

\bibitem{Fei2018}
Fei Z, Huang B, Malinowski P, Wang W, Song T, Sanchez J, Yao W, Xiao D, Zhu X,
  May A~F, Wu W, Cobden D~H, Chu J~H and Xu X 2018 {\em Nat. Mater\/} {\bf 17}
  778 

\bibitem{Deng2018}
Deng Y, Yu Y, Song Y, Zhang J, Wang N~Z, Sun Z, Yi Y, Wu Y~Z, Wu S, Zhu J, Wang
  J, Chen X~H and Zhang Y 2018 {\em Nature\/} {\bf 563} 94
  

\bibitem{May2019}
May A~F, Bridges C~A and McGuire M~A 2019 {\em Phys. Rev. Materials\/} {\bf
  3}(10) 104401

\bibitem{Tian2020}
Tian C, Pan F, Xu S, Ai K, Xia T and Cheng P 2020 {\em Appl. Phys. Lett.\/}
  {\bf 116} 202402

\bibitem{May2020}
May A~F, Du M~H, Cooper V~R and McGuire M~A 2020 {\em Phys. Rev. Materials\/}
  {\bf 4}(7) 074008
  

\bibitem{Ding2020}
Ding B, Li Z, Xu G, Li H, Hou Z, Liu E, Xi X, Xu F, Yao Y and Wang W 2020 {\em
  Nano Lett.\/} {\bf 20} 868--873 pMID: 31869236
  

\bibitem{wang2020characteristics}
Wang H, Wang C, Li Z~A, Tian H, Shi Y, Yang H and Li J 2020 {\em Appl. Phys.
  Letters\/} {\bf 116} 192403

\bibitem{wu2020neel}
Wu Y, Zhang S, Zhang J, Wang W, Zhu Y~L, Hu J, Yin G, Wong K, Fang C, Wan C
  {\em et~al.\/} 2020 {\em Nat. Commun\/} {\bf 11} 3860

\bibitem{yang2020creation}
Yang M, Li Q, Chopdekar R, Dhall R, Turner J, Carlstr{\"o}m J, Ophus C, Klewe
  C, Shafer P, N¡¯Diaye A {\em et~al.\/} 2020 {\em Sci. Adv.\/} {\bf 6}
  eabb5157

\bibitem{Park2021skrymion}
Park T~E, Peng L, Liang J, Hallal A, Yasin F~S, Zhang X, Song K~M, Kim S~J, Kim
  K, Weigand M, Sch\"utz G, Finizio S, Raabe J, Garcia K, Xia J, Zhou Y, Ezawa
  M, Liu X, Chang J, Koo H~C, Kim Y~D, Chshiev M, Fert A, Yang H, Yu X and Woo
  S 2021 {\em Phys. Rev. B\/} {\bf 103}(10) 104410
  

\bibitem{Wu2021}
Wu X, Lei L, Yin Q, Zhao N~N, Li M, Wang Z, Liu Q, Song W, Ma H, Ding P, Cheng
  Z, Liu K, Lei H and Wang S 2021 {\em Phys. Rev. B\/} {\bf 104}(16) 165101
  

\bibitem{Yang2021}
Yang X, Zhou X, Feng W and Yao Y 2021 {\em Phys. Rev. B\/} {\bf 104}(10) 104427
  

\bibitem{Yamagami2021}
Yamagami K, Fujisawa Y, Driesen B, Hsu C~H, Kawaguchi K, Tanaka H, Kondo T,
  Zhang Y, Wadati H, Araki K, Takeda T, Takeda Y, Muro T, Chuang F~C, Niimi Y,
  Kuroda K, Kobayashi M and Okada Y 2021 {\em Phys. Rev. B\/} {\bf 103}(6)
  L060403

\bibitem{Li2020magnetic}
Li Z, Xia W, Su H, Yu Z, Fu Y, Chen L, Wang X, Yu N, Zou Z and Guo Y 2020 {\em
  Sci. Reports\/} {\bf 10} 1--10

\bibitem{Tan2021gate}
Tan C, Xie W~Q, Zheng G, Aloufi N, Albarakati S, Algarni M, Li J, Partridge J,
  Culcer D, Wang X {\em et~al.\/} 2021 {\em Nano Lett.\/} {\bf 21} 5599--5605

\bibitem{Ly2021direct}
Ly T~T, Park J, Kim K, Ahn H~B, Lee N~J, Kim K, Park T~E, Duvjir G, Lam N~H,
  Jang K {\em et~al.\/} 2021 {\em Adv. Funct. Mater\/} {\bf 31} 2009758

\bibitem{Ohta2021butterfly}
Ohta T, Tokuda M, Iwakiri S, Sakai K, Driesen B, Okada Y, Kobayashi K and Niimi
  Y 2021 {\em AIP Advances\/} {\bf 11} 025014

\bibitem{Zhang2020}
Zhang H, Chen R, Zhai K, Chen X, Caretta L, Huang X, Chopdekar R~V, Cao J, Sun
  J, Yao J, Birgeneau R and Ramesh R 2020 {\em Phys. Rev. B\/} {\bf 102}(6)
  064417 

\bibitem{Seo2020}
Seo J, Kim D~Y, An E~S, Kim K, Kim G~Y, Hwang S~Y, Kim D~W, Jang B~G, Kim H,
  Eom G {\em et~al.\/} 2020 {\em Sci. Adv.\/} {\bf 6} eaay8912

\bibitem{Ohta2020}
Ohta T, Sakai K, Taniguchi H, Driesen B, Okada Y, Kobayashi K and Niimi Y 2020
  {\em Appl. Phys. Express\/} {\bf 13} 043005

\bibitem{Drachuck2018}
Drachuck G, Salman Z, Masters M~W, Taufour V, Lamichhane T~N, Lin Q, Straszheim
  W~E, Bud'ko S~L and Canfield P~C 2018 {\em Phys. Rev. B\/} {\bf 98}(14)
  144434 

\bibitem{Tian2019}
Tian C~K, Wang C, Ji W, Wang J~C, Xia T~L, Wang L, Liu J~J, Zhang H~X and Cheng
  P 2019 {\em Phys. Rev. B\/} {\bf 99}(18) 184428
  

\bibitem{Verchenko2015}
Verchenko V, Tsirlin A, Sobolev A, Presniakov I and Shevelkov A 2015 {\em
  Inorg. Mater.\/} {\bf 54} 8598

\bibitem{May2016}
May A~F, Calder S, Cantoni C, Cao H and McGuire M~A 2016 {\em Phys. Rev. B\/}
  {\bf 93}(1) 014411
  

\bibitem{Yuan2017}
Yuan D, Jin S, Liu N, Shen S, Lin Z, Li K and Chen X 2017 {\em Mater. Res.
  Express\/} {\bf 4} 036103

\bibitem{Supporting}
 {\em supporting materials\/}

\bibitem{May2020practical}
May A~F, Yan J and McGuire M~A 2020 {\em Journal of Applied Physics\/} {\bf
  128} 051101

\bibitem{Gao2020}
Gao Y, Yin Q, Wang Q, Li Z, Cai J, Zhao T, Lei H, Wang S, Zhang Y and Shen B
  2020 {\em Adv. Mater\/} {\bf 32} 2005228

\bibitem{Verchenko2016new}
Verchenko V~Y, Sokolov S~S, Tsirlin A~A, Sobolev A~V, Presniakov I~A, Bykov
  M~A, Kirsanova M~A and Shevelkov A~V 2016 {\em Dalton Trans.\/} {\bf 45}
  16938--16947

\bibitem{FullProf}
Rodriguez-Carvajal J 1993 {\em Physica B\/} {\bf 192} 55--69

\bibitem{Sheldrick2015}
Sheldrick G~M 2015 {\em Acta Crystallogr. C Struct. Chem.\/} {\bf C71} 3

\bibitem{PerdewGGA1996}
Perdew J~P, Burke K and Ernzerhof M 1996 {\em Phys. Rev. Lett.\/} {\bf 77}(18)
  3865--3868

\bibitem{Kresse1996a}
Kresse G and Furthm{\"{u}}ller J 1996 {\em Computational Materials Science\/}
  {\bf 6} 15

\bibitem{Kresse1999}
Kresse G and Joubert D 1999 {\em Phys. Rev. B\/} {\bf 59}(3) 1758--1775
  

\end{thebibliography}

\begin{thebibliography}{10}
\bibitem{May2020}
May A~F, Du M~H, Cooper V~R and McGuire M~A {\em Phys. Rev. Materials\/}
  {\bf 4}(7) 074008 (2020).
\end{thebibliography}


\end{document}